\documentclass{aa}  

\usepackage{graphicx}
\usepackage{txfonts}
\usepackage{natbib}
\usepackage{mathrsfs}
\usepackage{longtable}
\usepackage[shortlabels]{enumitem}
\usepackage{siunitx}
\usepackage{epstopdf}
\usepackage{hyperref}   % Hyperlinks
\hypersetup{colorlinks=true,linkcolor=blue,citecolor=blue,filecolor=blue,urlcolor=blue}
\usepackage{comment}
\usepackage{txfonts}
\usepackage{url}
\usepackage{multirow} 
\usepackage{ulem}
\begin{document} 

%\title{Nuclear High-Ionization Outflow in the Compton-Thick AGN NGC~6552 as seen with \textit{JWST} Mid-Infrared Spectroscopy}
\title{Nuclear high-ionisation outflow in the Compton-thick AGN NGC~6552 as seen by the \textit{JWST} mid-infrared instrument}

   %\subtitle{Characterizing the MRS Resolving Power}
   \author{ 
J. \'Alvarez-M\'arquez\inst{\ref{cab}}\thanks {Email:javier.alvarez@cab.inta-csic.es}, %\orcidlink{0000-0002-7093-1877} 
A.\ Labiano\inst{\ref{tpz},\ref{cab}}, %\orcidlink{0000-0002-0690-8824}
P.\ Guillard\inst{\ref{sorb},\ref{iuf}}, %\orcidlink{0000-0002-2421-1350}
D.\ Dicken\inst{\ref{ukatc}}, %\orcidlink{0000-0003-0589-5969}
I.\ Argyriou\inst{\ref{kul}}, %\orcidlink{0000-0003-2820-1077}
P.\ Patapis\inst{\ref{eth}}, %\orcidlink{0000-0001-8718-3732}
D.~R.\ Law\inst{\ref{stsci}}, %\orcidlink{0000-0002-9402-186}
P. J. Kavanagh\inst{\ref{dias}}, %\orcidlink{0000-0001-6872-2358}
K.~L.\ Larson\inst{\ref{aura}}, %\orcidlink{0000-0003-3917-6460}
D.\ Gasman\inst{\ref{kul}}, %\orcidlink{0000-0002-1257-7742}
M.\ Mueller\inst{\ref{kai},\ref{sron}}, %\orcidlink{0000-0003-3217-5385}
S.\ Alberts\inst{\ref{Arizona}}, %\orcidlink{0000-0002-8909-8782}
B.~R.\ Brandl\inst{\ref{leid},\ref{delft}}, %\orcidlink{0000-0001-9737-169X}
L. Colina\inst{\ref{cab}}, %\orcidlink{0000-0002-9090-4227}
M. Garc\'ia-Mar\'in\inst{\ref{esa-st}},
O.~C.\ Jones \inst{\ref{ukatc}}, %\orcidlink{0000-0003-4870-5547}
A.\ Noriega-Crespo\inst{\ref{stsci}}, %\orcidlink{0000-0002-6296-8960}
I.\ Shivaei\inst{\ref{Arizona}}, %\orcidlink{0000-0003-4702-7561}
T.\ Temim\inst{\ref{pu}}, %\orcidlink{0000-0001-7380-3144}
G.~S. Wright\inst{\ref{ukatc}}
%A.~M. Glauser\orcidlink{0000-0001-9250-1547}\inst{\ref{eth}},
%O.~D.\ Fox\orcidlink{0000-0003-2238-1572}\inst{\ref{stsci}},
%B.\ Sargent\orcidlink{0000-0003-1689-9201}\inst{\ref{stsci},\ref{jhu}},
}

\institute{Centro de Astrobiolog\'ia (CAB), CSIC-INTA, Carretera de Ajalvir km4, 28850 Torrej\'on de Ardoz, Madrid, Spain. \label{cab}
\and Telespazio UK for the European Space Agency (ESA), ESAC, Camino Bajo del Castillo s/n, 28692 Villanueva de la Ca\~nada, Spain. \label{tpz}
\and Sorbonne Universit\'e, CNRS, UMR 7095, Institut d’Astrophysique de Paris, 98bis Bd Arago, 75014 Paris, France. \label{sorb}
\and Institut Universitaire de France, Minist\`ere de l’Enseignement Supérieur et de la Recherche, 1 rue Descartes, 75231 Paris Cedex 05, France. \label{iuf}
\and UK Astronomy Technology Centre, Royal Observatory Edinburgh, Blackford Hill, Edinburgh EH9 3HJ, UK \label{ukatc}
\and Institute of Astronomy, KU Leuven, Celestijnenlaan 200D bus 2401, 3001 Leuven, Belgium \label{kul}
\and Institute for Particle Physics and Astrophysics, ETH Zurich, Wolfgang-Pauli-Str 27, 8093 Zurich, Switzerland \label{eth}
\and Space Telescope Science Institute, 3700 San Martin Drive, Baltimore, MD 21218, USA \label{stsci}
\and Dublin Institute for Advanced Studies, Astronomy \& Astrophysics Section, 31 Fitzwilliam Place, Dublin 2, Ireland \label{dias}
\and AURA for the European Space Agency, Space Telescope Science Institute, 3700 San Martin Drive, Baltimore, MD 21218, USA \label{aura}
\and Kapteyn Astronomical Institute, University of Groningen, P.O. Box 800, 9700 AV Groningen, The Netherlands \label{kai}
\and SRON Netherlands Institute for Space Research, Postbus 800, 9700 AV Groningen, The Netherlands \label{sron}
\and Steward Observatory, University of Arizona, 933 N. Cherry Tucson, AZ 85721, USA \label{Arizona}
\and Leiden Observatory, Leiden University, P.O. Box 9513, 2300 RA Leiden, The Netherlands \label{leid}
\and Faculty of Aerospace Engineering, Delft University of Technology, Kluyverweg 1, 2629 HS Delft, The Netherlands \label{delft}
\and European Space Agency, Space Telescope Science Institute, Baltimore, Maryland, USA \label{esa-st}
\and Princeton University, 4 Ivy Ln, Princeton, NJ 08544 \label{pu}
}

   \date{Received ...; accepted ...}

% \abstract{}{}{}{}{} 
% 5 {} token are mandatory
\newcommand{\mum}{$\mu$m}
\newcommand{\kms}{km s$^{-1}$}
\newcommand{\slice}{$_\mathrm{slice}$}
\newcommand{\mone}{$^{-1}$}
\newcommand{\mtwo}{$^{-2}$}
\newcommand{\isoa}{iso-$\alpha$}
\newcommand{\isol}{iso-$\lambda$}
\newcommand{\Lsun}{\mbox{L$_{\odot}$}}
\newcommand{\Msun}{\mbox{M$_{\odot}$}}
\newcommand{\Myr}{\mbox{M$_{\odot}$ yr$^{-1}$}}
\newcommand{\HI}{\mbox{H\,{\sc i}}}
\newcommand{\HII}{\mbox{H\,{\sc ii}}}
\newcommand{\Hmol}{\mbox{H$_{\rm 2}$}}
\newcommand{\nH}{$n_{\rm H}$}
\newcommand{\Wm}{W~m$^{-2}$}

  \abstract
   {During the commissioning of the \textit{James Webb Space Telescope} (\textit{JWST}), the mid-infrared instrument (MIRI) observed NGC~6552 with the MIRI Imager and the Medium-Resolution Spectrograph (MRS). NGC~6552 is an active galactic nucleus (AGN) at a redshift of 0.0266 (D$_{L}$ = 120 Mpc) classified as a Seyfert 2 nucleus in the optical and Compton-thick AGN in the X-ray.}
   {This work exemplifies and demonstrates the MRS capabilities to study the mid-infrared (mid-IR) spectra and characterise the physical conditions and kinematics of the ionised and molecular gas in the nuclear regions of nearby galaxies.}
   {MIRI Imager observations covers the full NGC~6552 galaxy at 5.6$\mu$m. MRS observations covers its nuclear region (3.6$\times$4.3\,kpc at 17.7-27.9\,$\mu$m) in a wavelength range between 4.9 to 27.9 $\mu$m. These observations were obtained with the aim to investigate the persistence of the MIRI detectors (residual signal left from previous bright source observations). However, NGC~6552 observations demonstrate the performance and power of the MIRI instrument even with a non-optimal observational strategy.}
   {We obtained the nuclear, circumnuclear, and central mid-IR spectra of NGC~6552. They provide the first clear observational evidence for a nuclear outflow in NGC~6552. The outflow contributes to 67$\pm$7\% of the total line flux independent of the ionisation potential (27 to 187 eV) and critical densities (10$^4$ to 4$\times$10$^{6}$ cm$^{-3}$), showing an average blue-shifted peak velocity of -127$\pm$45 km\,s$^{-1}$ and an outflow maximal velocity of 698$\pm$80 km\,s$^{-1}$. Since the mid-IR photons penetrate dusty regions as efficiently as X-ray keV photons, we interpret these results as the evidence for a highly ionised, non-stratified, AGN-powered, and fast outflowing gas in a low density environment (few 10$^{3}$ cm$^{-3}$) located very close (<0.2\,kpc) to the Compton-thick AGN. Nine pure rotational molecular Hydrogen lines are detected and spectrally resolved, and exhibit symmetric Gaussian profiles, consistent with the galactic rotation, and with no evidence of outflowing \Hmol\ material. We detect a warm \Hmol\ mass of $1.9 \pm 1.1 \times 10^7~\Msun$ in the central region (1.8~kpc in diameter) of the galaxy, with almost 30\% of that mass in the circumnuclear region. Line ratios confirm that NGC~6552 has a Seyfert nucleus with a black hole mass estimated in the range of 0.6 to 6 million solar masses.}
   {This work demonstrates the power of the newly commissioned MIRI Medium Resolution Spectrograph to reveal new insights in the kinematics and ionisation state of the interstellar medium around the dusty nuclear regions of nearby active galaxies.}

    \titlerunning{MIRI observations of NGC~6552}

    \authorrunning{J. \'Alvarez-M\'arquez et al.}

   \maketitle
izi%-------------------------------------------------------------------

\section{Introduction}\label{intro}

\begin{table*}[!ht]
\caption{Summary of MIRI commissioning observations for NGC6552}
%\centering
\begin{tabular}{cccccccccc}
\hline
Visit & Imager target & MRS target & Imager Filter & MRS band & Groups & Int. & Exp. & Dithers & Exp. time [s] \\
 (1) & (2) & (3) & (4) & (5) & (6) & (7) & (8) & (9) & (10) \\
\hline
1 & Back. & Star & F560W  & SHORT & 25 & 1 & 1 & 4  & 277.5 \\
2 & NGC~6552 & Back. & F560W  & SHORT & 200 & 1 & 1 & no dither  & 555 \\
3 & NGC~6552 & Back. & F560W  & MEDIUM & 25 & 8 & 1 & no dither  &  574\\
4 & NGC~6552 & Back. & F560W  & LONG & 25 & 1 & 1 & 8  &  555\\
5 & Back. + Persistence & NGC~6552 & F560W  & ALL & 182 & 1 & 1 & no dither  & $3 \times 505$ \\
6.1 & Back. + Persistence & Back. & F560W  & SHORT & 25 & 1 & 1 & 4   & 277.5 \\
6.2 &  Back. + Persistence & Back.  & F2550W & SHORT & 25 & 1 & 1 & 4  & 277.5 \\
6.3 & Back. + Persistence & Back. & F560W  & SHORT & 25 & 20 & 1 & no dither  & 1440.2 \\
\hline
\end{tabular}
\label{tab:table-obs}
\tablefoot{Columns describe the visit number of the observation (1), the pointing target for the imager (2) and MRS (3), the filter used in the Imager (4), the band configuration used in the MRS (5), the number of groups (6), integrations (7), exposures (8), and dithers (9) for each observation, as well as the exposure time calculated as 2.77s x Groups x Int. x Exp. x Dithers (10). Consult PID1039 for details.}
\end{table*}

The Mid-Infrared Instrument \citep[MIRI,][]{miri_pasp_1,Wright+15} on board the \textit{James Webb Space Telescope} (\textit{JWST}) includes imaging and medium-resolution integral field spectroscopy (IFS) observing modes from 4.9 to 27.9 $\mu$m. The MIRI Imager provides sub-arcsec resolution (0.2-0.8 arcsec) images in nine filters, with a clear field of view (FoV) up to 2.3 square arcmin \citep{Bouchet15}.
%, which corresponds to $\sim$2500 kpc$^2$ at the distance of NGC6552. 
The Medium Resolution Spectrometer (MRS) provides IFS in twelve individual spectral bands, organised in four channels (1, 2, 3, and 4) and each of them is divided in three bands (SHORT, MEDIUM, and LONG). The MRS provides a factor of x10-100 increased sensitivity from previous mid-IR telescopes, combined with a sub-arcsec resolution (0.3 - 0.9 arcsec) in a FoV ranging from $\sim$13 to $\sim$56 square arcsec, and a spectral resolution of 75 to 200 km\,s$^{-1}$ (\citealt{Wells15, Glasse+15, Labiano21,Jones+23}, Argyriou et al. in prep.). Due to the low central extinction in the mid-IR spectral range, similar to that in the several keV X-ray range \citep{Corrales2016}, and the improvements in the spatial and spectral resolution from previous mid-IR instruments, MIRI will be a key instrument in providing a view into the inner dust-enshrouded nuclear regions of nearby galaxies and offering the possibility of separating the nucleus from its circumnuclear environment and studying both components with unprecedented detail and sensitivity. 

Galactic winds and outflows are an universal phenomenon associated with AGNs and star-forming galaxies \citep{Veilleux+05, Fabian2012, King+15, Veilleux+20}. They have been detected in different phases of the gas, from highly ionised to ionised, neutral, and molecular gas, and at both low-z and high-z \citep{Rodriguez-Ardila+06, Guillard+12, Veilleux+13, Arribas+14, Harrison+16, Emonts+17, Pereira-Santaella+18, Davies+20, Spilker+20, Alvarez-Marquez+21}. The mid-IR wavelength range is very rich in coronal emission lines, that is, lines  with ionisation potential (E$_{ion}^{*}$) that is close or above 100 eV, of different elements (sulfur, neon, iron, argon, magnesium, among others). These lines, together with the low extinction in this spectral range relative to the optical, provide a unique opportunity to probe the gas close to the AGN and therefore to establish the physical conditions and kinematics of the outflows in the inner regions close to the black hole and accreting torus \citep[e.g.,][]{Pereira-Santaella+10,Fischer11}.

NGC~6552 is a galaxy located at a luminosity distance (D$_{L}$) of 120 Mpc (z=0.02656$\pm$0.00042) that is classified as an active galactic nucleus (AGN) Seyfert 2 in the optical \citep{Moran1996, Bower96, Falco99, 2002LEDA, Gioia03, Shu07}, as well as a Compton-thick AGN (CT-AGN), with a derived column density of $\log(N_{H})$ = 24.05 cm$^{-3}$, in the X-ray \citep{Ricci2015}. The most recent X-ray studies with \textit{XMM-Newton} \citep{Jansen+01}, complemented with \textit{NuSTAR} \citep{Harrison+13} observations sensitive to hard X-rays in the 3 to 78 keV range, have confirmed the classification of NGC~6552 as a CT-AGN \citep{Torres-Alba_21}. The X-ray model predicts the presence of a CT-AGN within a Compton-thin torus oriented nearly edge-on, at 75 degrees with respect to our line of sight \citep{Torres-Alba_21}. NGC~6552 has also been observed in imaging and spectroscopy at mid-infrared (mid-IR) wavelengths with \textit{Spitzer} and \textit{WISE}. These observations have been used to cross-calibrate the long-wavelength band of \textit{WISE} (W4) with the \textit{Spitzer} InfraRed Spectrograph long-low module (IRS-LL) and MIPS 24$\mu$m channel \citep{Jarrett11}.

In this paper, we demonstrate the MIRI capabilities to study the physical conditions and kinematics of the nuclear regions in nearby galaxies, with the example of NGC 6552. The paper is structured as follows: Section \ref{obs.sec} describes the MIRI Imager and MRS observations of NGC~6552. Sections \ref{MIRI.imager} and \ref{Cal.sec} explains the calibration of the MIRI Imager and MRS data. Section \ref{Spectra.sec} shows the final calibrated spectra of NGC~6552. Section \ref{disc.sec} presents the fluxes and kinematics measurements of all detected emission lines, the properties of the ionised outflowing gas and the warm molecular hydrogen, and the Seyfert nature of the AGN and black hole mass of NGC~6552. Finally, Section \ref{sum.con.sec} summarises and concludes the paper. We have adopted $H_{0} = 67.7$ km s$^{-1}$ Mpc$^{-1}$ and $\Omega_{\rm m} = 0.307$ \citep{Cosmology2016} as cosmological parameters. The corresponding luminosity distance (D$_{L}$) and physical scale used in this paper for z=0.02656 are 120 Mpc and 0.552 kpc\,arcsec$^{-1}$, respectively \citep{CosmoCalc}. 

\section{Observations}\label{obs.sec}

MIRI data of NGC6552 was obtained on May 7, 2022, during the commissioning of \textit{JWST}, as part of the persistence characterisation of the MIRI detectors (\href{https://www.stsci.edu/jwst/science-execution/program-information.html?id=1039}{PID 1039}, P.I. D. Dicken). NGC6552 was selected because is one of the brightest mid-IR source in the northern continuous viewing zone of JWST. Aditionaly, NGC6552 was used as a calibrator of WISE and \textit{Spitzer,} offering the possibility of cross-checking the MRS absolute photometry. 

\begin{figure*}[!ht]
\begin{center}
\includegraphics[width=\hsize]{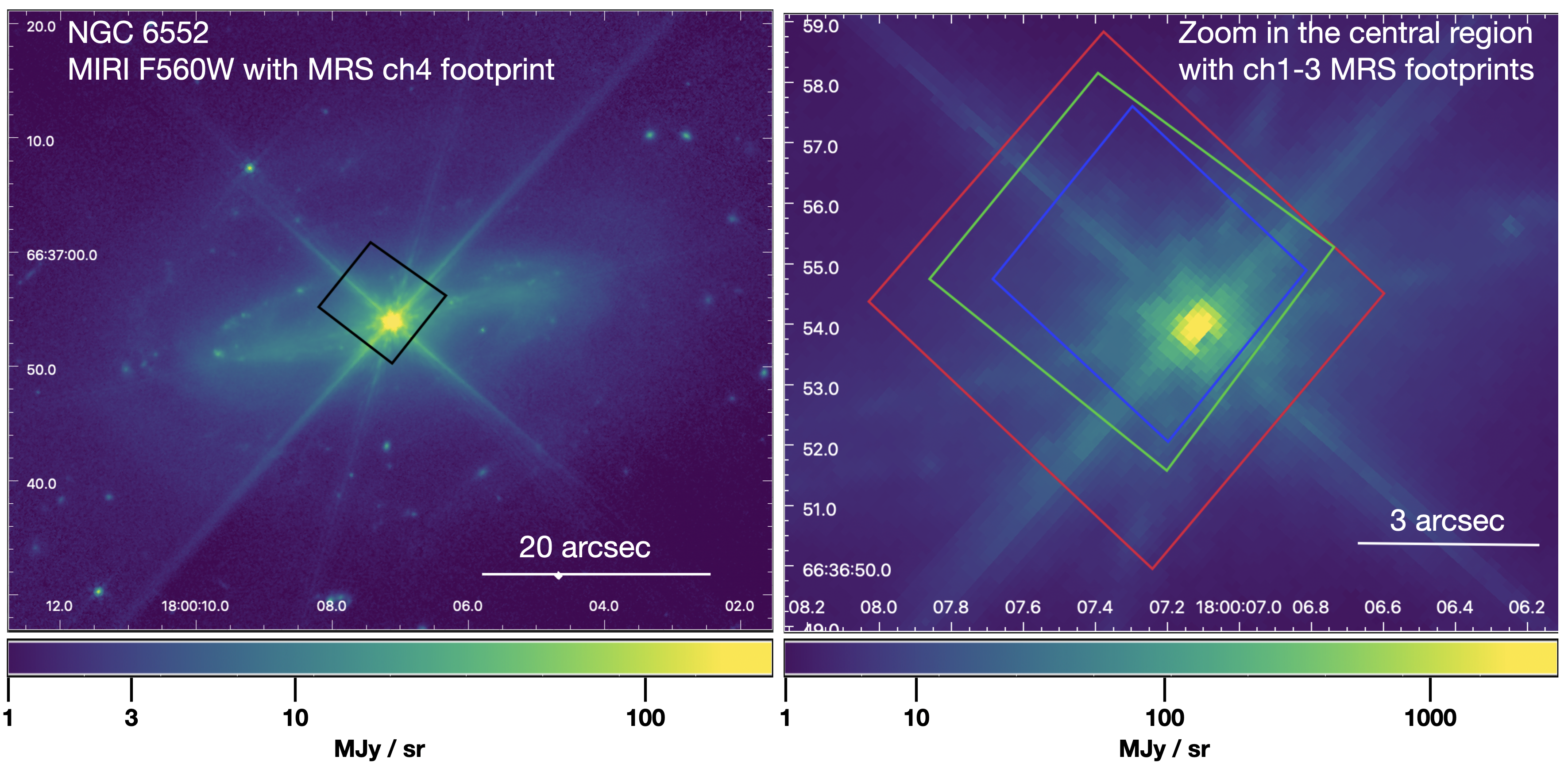}
\caption{MIRI background-subtracted images of NGC~6552 in the F560W filter. Left panel shows the entire galaxy (FoV$\sim$20x20 kpc) and the right panel is a zoom on the central region (FoV$\sim$4x4 kpc) observed with the MRS. The rectangles show an overlay of the four MRS channels (from the short to long wavelengths: CH1 (FoV=2x2 pc$^2$) in blue, CH2 (FoV=2.6x2.5 pc$^2$) in green, CH3 (FoV=3.4x3.4 pc$^2$) in red on the zoom-in right panel, and CH4 (FoV=4x4 pc$^2$) in black on the left panel. Both images are shown on a log-scale, but with different stretches and contrasts to highlight the extended faint ring structure in the left panel, and the point-like central active nuclei in the right panel. The dynamical range in flux between the central pixels and the extended ring is about 20000, showing the exquisite imaging capabilities of the MIRI detectors.} 
%The bottom panels show comparisons with the Spitzer IRAC 3.6$\mu$m (left) and the HST/ACS F814W (right) images, over a similar field of view.}
\label{fig:images}
\end{center}
\end{figure*}

The aim of this program was to investigate aspects of persistence after long observations of NGC 6552 on the  MIRI Imager and MRS detectors and to analyze any artefacts and their decay for up to 25 minutes. The magnitude of the detector persistence is a function of flux of the source of origin, as well as the time spent observing that source on the same detector pixel \citep{Rieke+15}. Such persistent behaviour was  first measured during MIRI ground testing, so this commissioning activity aimed at verifying that the memory effect seen in the detectors after observing a bright source was as expected.

%The observations were taken using the simultaneous mode of MIRI, where the Imager and the MRS observe simultaneously slightly different field of views. 
The program began by taking a background observation with the filter F560W of the MIRI Imager and the MRS observing a star (TYC-4213-1049-1), using the SHORT band (visit 1). The aim of this observation was to gauge the state of the detectors (i.e. to look for any existing artefacts) before the programme started, as well as to use it for background subtraction in the MIRI imager calibration process. A carefully designed sequence of observations followed, using both the MIRI Imager and the MRS simultaneously observing adjacent FoV (see Table~\ref{tab:table-obs} for a summary). Visits 2, 3, and 4 were relatively long (555 seconds) on-target exposures with the MIRI imager aimed at putting a large amount of signal on the detector to see if this imprints a persistence signature on the data. The nucleus of NGC~6552 was able to saturate the imaging data in the third detector readout. Visit 5 was on-target MRS observations with exposure times of 505 seconds in each of the MRS bands. The MRS observations did not saturate. Visit 6 was used as a reference image and to study persistence decay in the MIRI Imager detector, if any. Finally, MRS observations from visits 2, 3, and 4 have been used for background characterisations purposes. 

Since the MRS is spatially undersampled, the MIRI instrument team recommends using four-point dithering\footnote{All the details on dithering patterns for the different MIRI modes of observation are available in the \href{https://jwst-docs.stsci.edu/jwst-mid-infrared-instrument/miri-operations/miri-dithering}{JDox MIRI dithering pages}} on every MIRI observation to maximise the spatial and spectral sampling and the scientific return of the MIRI observations. Due to the nature of this program, the MRS on-target observations did not use dithers, as the programme was designed to investigate artefacts in the detector image rather than on sky. The MIRI Imager visits 2 and 3 used offsets instead of dithers to make sure any persistence artefacts were well separated from one image to another. The MIRI Imager visit 4 uses two four-point, extended dither patterns. The MRS background simultaneous observations (visits 2, 3, and 4) follow the MIRI Imager dither pattern strategies. For additional information on dithers and exposure times see Table~\ref{tab:table-obs}.

The results of the commissioning investigation demonstrated that the impact of persistence artefacts was very low, where persistence was seen at a level <0.01\% of the source flux. Such levels of persistence are easily calibrated using standard MIRI dither patterns and the JWST calibration pipeline. In the case of NGC~6552, the saturating nucleus had no impact on the imaging or subsequent exposures, as the MIRI Imager proved to work extremely well in this relatively high-contrast observation (see Section \ref{MIRI.imager}).

\section{MIRI image}\label{MIRI.imager}

The MIRI imaging data were reprocessed with the v1.6.3 \textit{JWST} pipeline, using the latest reference files available\footnote{the version of CRDS was 11.16.6, and CRDS context "jwst\_0942.pmap"}. After the level 1 detector processing, we applied a fix to the pipeline to re-create rate images averaging only valid data, as a bug in the pipeline affected the pixels that are saturating in after the third group, which is the case for some of the central pixels of NGC6552. For those pixels, only the first three frames were used. We note that this fix will be implemented in a future version of the pipeline. Then, the level 2 pipeline was run with the default parameters and using the sky flats computed from the LMC data taken during commissioning (PID~1040). These preliminary flats contain image artifacts caused by MIRI reset anomaly \citep[see][for a discussion of this effects]{Ressler+15}, such as tree-ring shaped structures. However, most of these image artifacts are removed by the master background subtraction (explained below), as the features are stable.
Since the calibration of the image is still preliminary, the MIRI Imager data are only used in this paper to illustrate the MRS pointings, crossmatch the MIRI Imager and the MRS photometry, and illustrate the optical quality (in particular the low persistence) and dynamical range of the imager. 

A master sky background image was created from the first visit file and subtracted from all the images. This sky image was created by sigma clipping all the images from the visit 1 aligned in instrument coordinates. We note that this injects residual small scale variations at a percent level in the background, which are not critical for our analysis. The left panel Fig.~\ref{fig:images} shows on  the entire galaxy and the footprints of the MRS channel 4, highlighting the low surface brightness emission in the dusty ring around the barred galaxy. Clumps of star formation are resolved in the bar, the spiral arms and the ring. The right panel shows the central nucleus with the footprints of MRS channels 1, 2, and 3. The MRS footprints are not perfectly centred onto the nucleus because we did not use target acquisition and because those observations were done before updating the distortion model (how the light from the FoV travels through the instrument and gets projected on the detectors) with the flight data of the observatory. The nucleus image exhibits the bright central cross-shaped PSF that is characteristic of the cruciform pattern at this wavelength \citep{Gaspar+21}, which shows that the AGN is dominating the central point-like emission. 

\begin{figure}[!ht]
\begin{center}
 \includegraphics[scale=0.39]{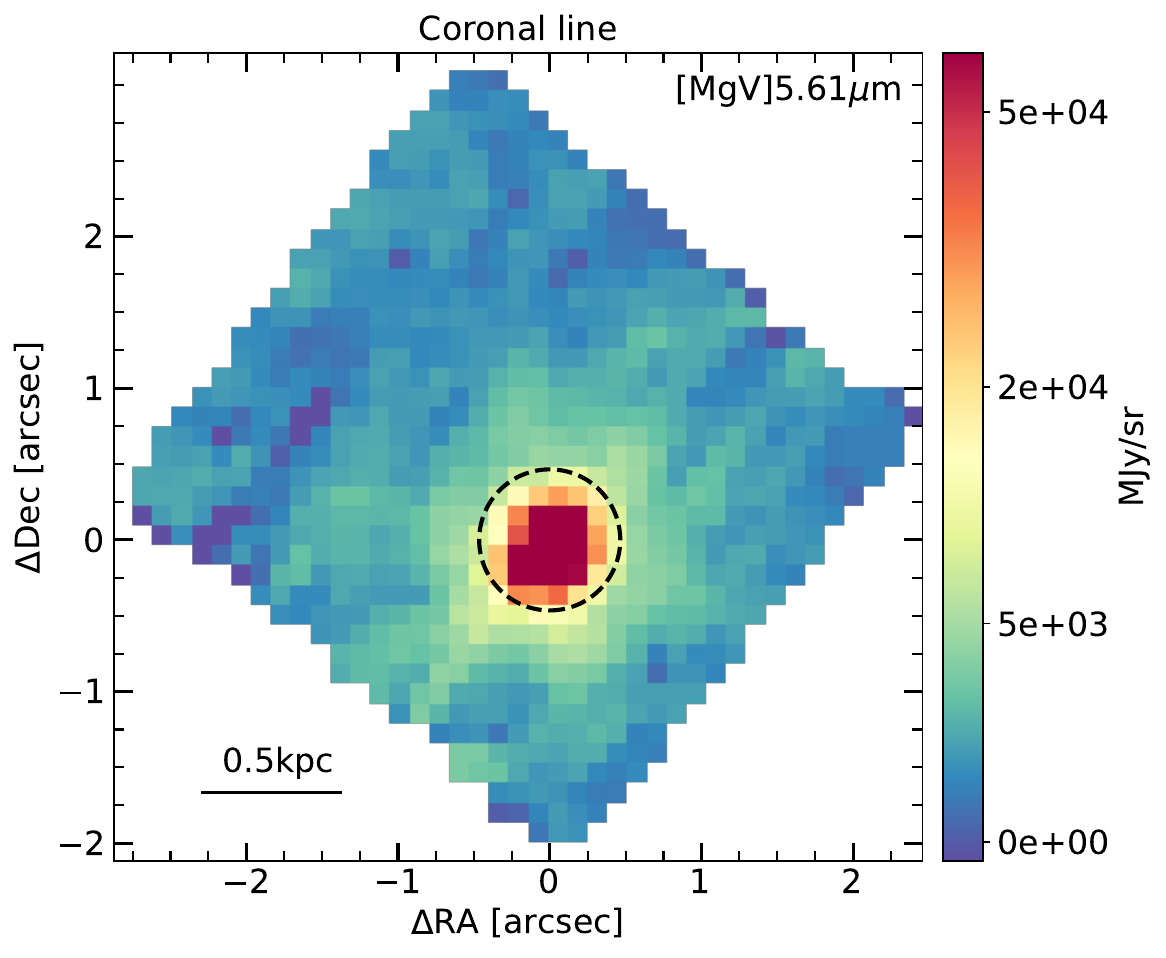} 
 \includegraphics[scale=0.39]{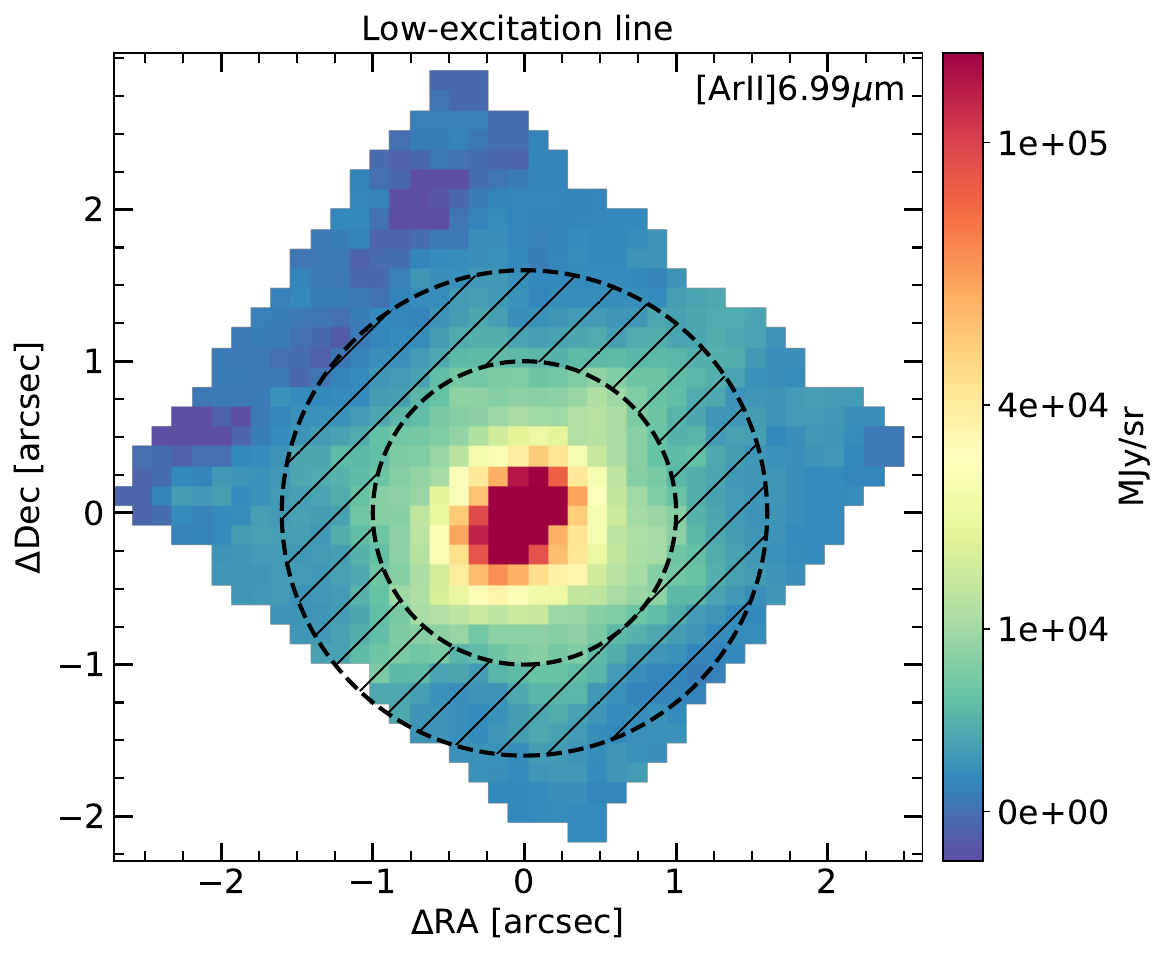}
 \includegraphics[scale=0.39]{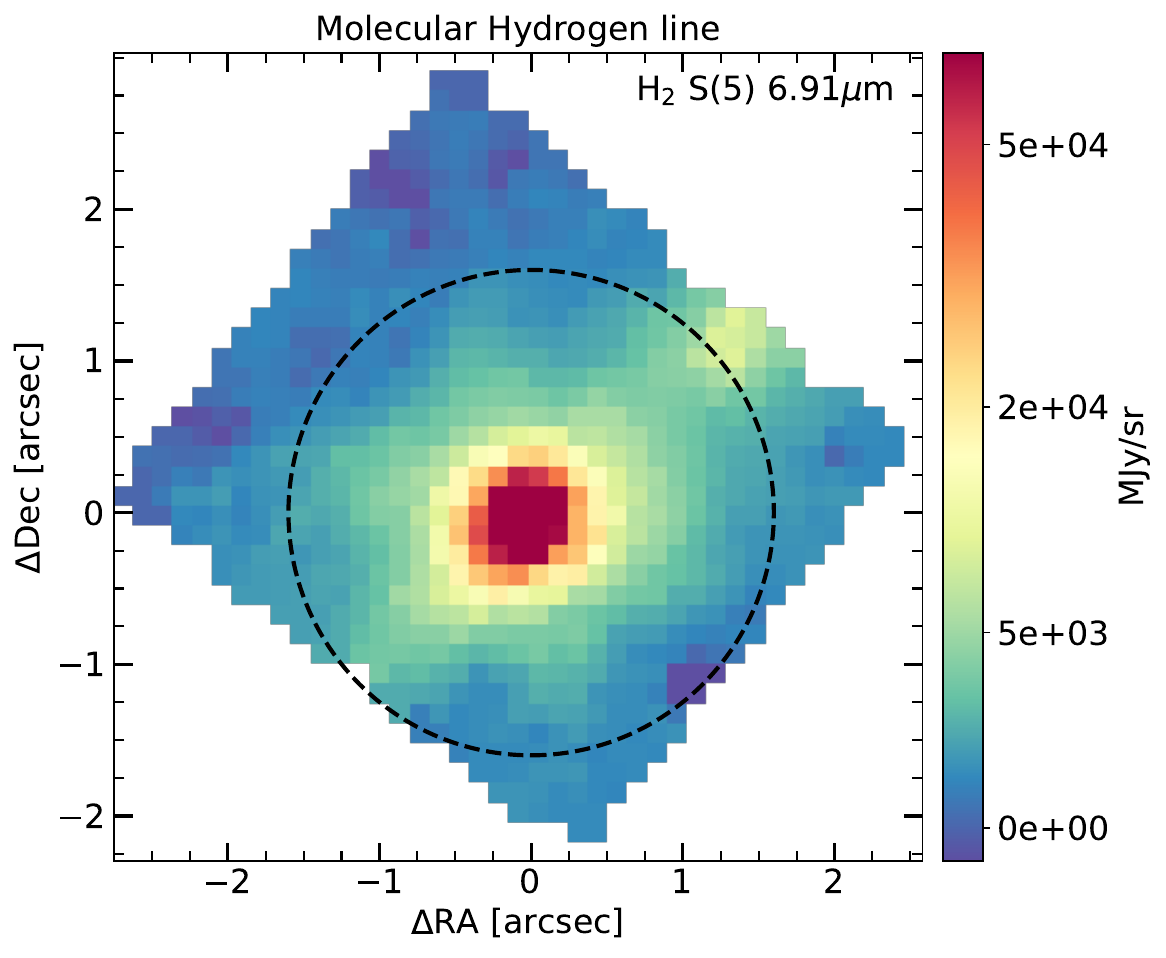}
\caption{Emission line maps in the central region of NGC6552 generated using the MRS channel 1 (3.2x3.7\,arcsec). Upper panel:\ Line map of [MgV]5.61$\mu$m as a representation of a coronal line emission, and illustrate the aperture used to extract the nuclear spectrum of NGC6552 in channel 1. Middle panel:\ Line map of [ArII]6.99$\mu$m as a representation of a low-excitation line emission, and illustrate the aperture used to extract the circumnuclear spectrum of NGC6552. Bottom panel:\ Line map of \Hmol(0-0)S(5) as a representation of the molecular hydrogen line emission and an illustration of the aperture used to extract the central spectrum of NGC6552.} \label{linemap.plot}
\end{center}
\label{maps}
\end{figure}

\section{MIRI Spectroscopy}
\subsection{MRS data calibration}\label{Cal.sec}

The MRS observations were processed with the \textit{JWST} calibration pipeline (release 1.8.3). This release uses the build 8.4.2 of the Data Management System (DMS) and context 1017 of the Calibration References Data System (CRDS). The MRS pipeline is divided into three different processing stages \citep{MRSpipeline,bushouse_howard_2022_6984366}\footnote{Check this  \href{https://jwst-docs.stsci.edu/jwst-science-calibration-pipeline-overview}{JDox page} for general information about the \textit{JWST} calibration pipeline. Review the \href{https://jwst-docs.stsci.edu/jwst-mid-infrared-instrument/miri-features-and-caveats}{MIRI Features and Caveats} and \href{https://jwst-docs.stsci.edu/jwst-calibration-pipeline-caveats}{Pipeline Caveats} webpages for the latest status, and information on the MIRI performance and calibration pipeline known issues.\label{Prueba_footnote}}. The first stage performs the detector-level corrections, identification, and correction of the cosmic ray (CR) impacts and then transforms the ramps\footnote{Check this \href{https://jwst-docs.stsci.edu/understanding-exposure-times}{link} for the ramp definition in JWST observations.} into slope detector products (Morrison et al. in prep). The second stage assigns the coordinate system, performs the straylight, fringe flat, residual fringe correction, and photometric calibrations to generate fully calibrated individual exposures (Argyriou et al. in prep.). The third stage combines the different dithered exposures to create 3D spectral cubes and 1D extracted spectra (Law et al. in prep). In the process, it allows for the performance of an outlier rejection and a background matching. The background could be subtracted in stage 2 or 3 of the pipeline depending of the methodology selected by the user. The third stage was not used in this work as the MIRI data was not dithered, and the 1D spectral extraction and background subtraction were performed outside of the JWST calibration pipeline. 

We found that the standard pipeline identifies and corrects most of the CR events, but the so-called \textit{CR} 'showers' and/or high energetic CRs still leave some relevant residual effects even using the large CRs events function. Also, the current in-flight darks and reset anomaly corrections  inject a vertical striping pattern in the slope detector data (see Footnote \ref{Prueba_footnote} for examples). Their strength depends on the length of the ramp, where long ramps ($>$~150-200 groups in FASTR1 readout mode, $\sim$500s) minimise this effect. These effects, along with others such as hot or bad detector pixels, are vastly reduced in four-point dithered observations allowing to obtain the optimal MRS spatial and spectral resolution. The MRS observations of NGC~6552 were taken with no dithers, as the goal of this commissioning activity was not to obtain a final optimised cube. Then, the 3D cubes could still have a small number of spaxels affected by the lack of data due to bad or hot pixels in the detector. These have been identified, masked and interpolated in the 1d extracted spectra presented in Sec. \ref{Spectra.sec}. We generated 12 3D cubes, that is, one for each of the MRS sub-channels, with spatial sampling of 0.13"~$\times$~0.13", 0.17"~$\times$~0.17", 0.20"~$\times$~0.20", and 0.35"~$\times$~0.35" for channels 1, 2, 3, and 4, respectively.

We extracted the 1D spectra using the 3D cubes and following the standard aperture or annulus photometry \citep{larry_bradley_2022_6825092}. We generated three 1D spectra from: (i) a nuclear region, (ii) a circumnuclear region, and (iii) a larger aperture that includes both the nuclear and circumnuclear regions (see Section \ref{Spectra.sec} for details). All the 1D spectra were derived individually for each of the 12 MRS sub-bands. The background is calculated in an annulus far away of the nuclear and circumnuclear emission of the source and subtracted in the 1D spectra. The extracted 1D spectra are corrected for residual fringing using a post-pipeline spectral-level correction which is a modified version of the detector-level correction available in the JWST calibration pipeline (\citealt{Gasman+2022}; Kavanagh et al., in prep.). The correction reduced the final fringe residuals to levels lower than 6\%, and a median level of 2-4\% \citep{Rigby22}. To remove the (minor) flux discontinuities between channels, we stitched the 12 sub-channels together, using the channel 3 bands as a reference. We found that the multiplicative stitching factors needed for each individual channel was lower than $\pm$5\% in all sub-channels.

\subsection{Spectral Extraction}\label{Spectra.sec}

\begin{figure*}[t]
\begin{center}
 \includegraphics[width=\hsize]{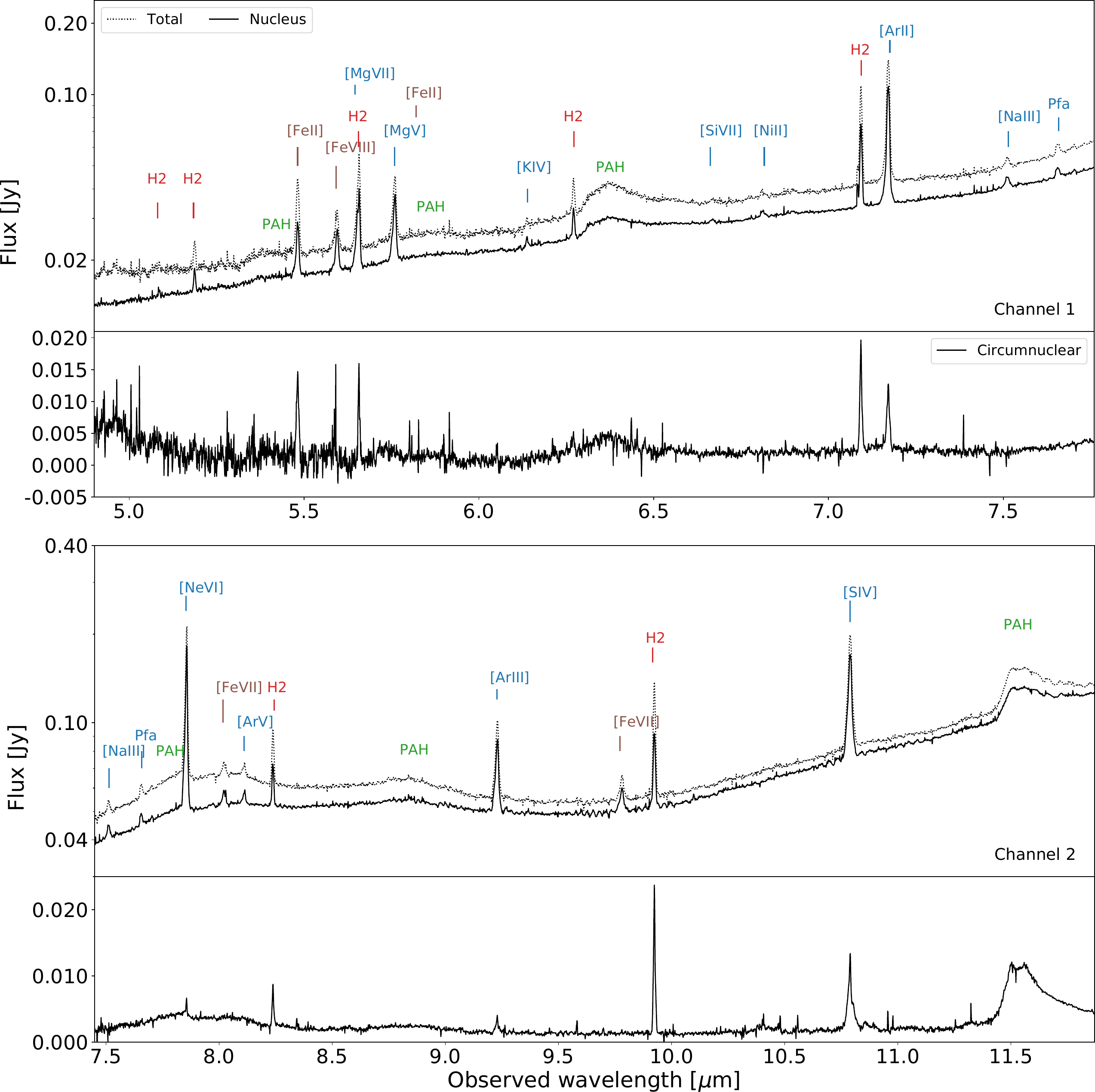} 
\caption{Nuclear, circumnuclear, and central MRS spectra of NGC~6552 from MRS channels 1 and 2. Upper panel:\ Central and nucleus spectra. Bottom:\ Circumnuclear spectra. The main emission features (listed in Table \ref{table.fluxes}) are highlighted together with the PAHs.} \label{spec1.plot}
\end{center}
\label{maps}
\end{figure*}

\begin{figure*}[t]
\begin{center}
 \includegraphics[width=\hsize]{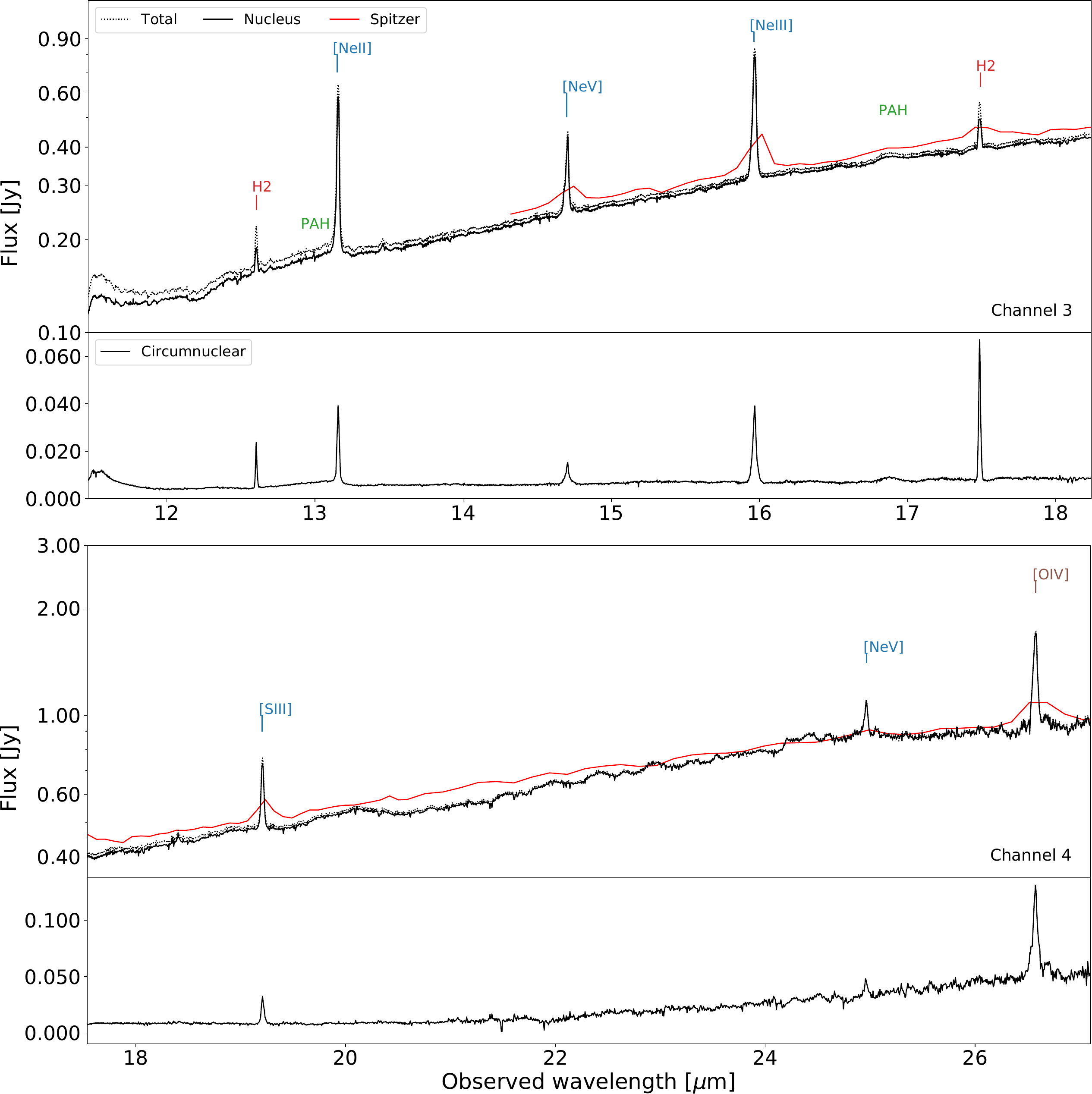}
\caption{As Fig. \ref{spec1.plot}, for MRS channels 3 and 4. The MRS channel 4 spectra is shown up to 27$\mu$m due to uncertainties in the photometrical calibrations for $\lambda > 27\mu$m. The comparison with the Spitzer IRS long low spectrum \citep[red line, from][]{Jarrett11} shows excellent photometric agreement from 14.5 to 22~\mum\ (see text for details) and illustrates how the gain in spectral resolution provided by the MRS helps to detect weak PAH features and weak lines (compared to the continuum), in particular, the high-excitation [NeV] lines. \label{spec2.plot}}
\end{center}
\label{maps}
\end{figure*}

As discussed in Section \ref{Cal.sec}, the calibrated 3D cubes of NGC~6552 are not optimally sampled because the MRS observations were not dithered. The analysis of this paper has focused in the characterisation of different areas of the central region of NGC~6552, rather than a more complex 2D analysis (e.g. \citealt{Pereira22}). Even so, the MRS provides an adequate angular resolution to separate the nuclear and circumnuclear regions in NGC~6552. If we assume D$_{L}$ = 120Mpc, the MRS point spread function (PSF) full width at half maximum (FWHM) corresponds to physical scales of 170$\,$pc for channel 1 evolving, up to a maximum of 580$\,$pc for channel 4LONG. As an example, Fig. \ref{linemap.plot} shows the central region of NGC~6552 observed by the MRS channel 1 (3.2x3.7\,arcsec). It presents three different line maps illustrating the emission from a coronal line (e.g. [MgV]5.61$\mu$m), a low-excitation line (e.g. [ArII]6.99$\mu$m), and a molecular hydrogen line (e.g. \Hmol(0-0)S(5)). They show a slightly different morphology between the different species. The coronal emission follows the MRS PSF, which is consistent with an emission coming form the nuclear region. Low-excitation emission is extending beyon the MRS PSF, consistent with the presence of low-excitation emission in the circumnuclear regions of NGC6552. The molecular hydrogen emission presents a similar morphology to the low-excitation  environment in the nuclear an circumnuclear region of NGC6552, but there is a secondary clump in the NW that is not present in the other species.

We performed three different 1D spectral extraction of the central region of NGC6552. First, we extracted a spectrum of the nuclear region using a circular aperture of radius equal to $1.5 \times FWHM (\lambda)$, where $FWHM (\lambda) = 0.3$ arcsec for $\lambda < 8\mu$m and $FWHM (\lambda) = 0.31\times \lambda[\mu m] / 8$ arcsec for $\lambda > 8\mu$m (see Fig. \ref{linemap.plot} upper panel). The selected FWHM\,($\lambda$) values follow the MRS PSF FWHM and it is the recommended aperture to extract a 1D spectrum of a point-like sources in MRS observations (Law et al. in prep.). We used the MRS PSF models (Patapis et al. in prep.) to correct the aperture losses in the 1D spectra. The percentage of flux that losses out of the selected aperture is 17\% for channel 1 and evolves up to 30\% in channel 4. Second, we extracted a spectrum of the circumnuclear region in an annulus with an inner radius equal to 1 arcsec, $\sim$0.55 kpc, and an outer radius of 1.6 arcsec, $\sim$0.88 kpc (see Fig. \ref{linemap.plot} middle panel). The inner radius is selected to fit the channel 4 MRS PSF FWHM, which is the minimum distance that allows for the disentanglement of the nuclear and circumnuclear emission in all MRS channels. The outer radius is delimited by the largest aperture allowed by the MRS FoV of channel 1. Based on the MRS PSF models and assuming the nucleus as a point-like source, the circumnuclear region is contaminated by a 2.5\%, in channel 1, and up to 12\%, in channel 4, of flux coming from the nucleus. We corrected the circumnuclear spectrum from the nuclear contamination by subtracting the percentage of the nuclear emission enclosed in the selected annulus aperture using the nuclear spectrum. Third, we extracted a spectrum (named 'central') in a circular aperture with a constant radius of 1.6 arcsec, $\sim$0.88 kpc, for all MRS channels (see Fig. \ref{linemap.plot} bottom panel). The selected radius provides the largest circular aperture allowed by the FoV of the MRS channel 1. Assuming a nuclear emission as a point-like source, a fraction of the nuclear emission will be extended outside of the selected aperture. The percentage of nuclear flux that is lost out of the selected aperture ranges from 6\%, in channel 1 to 20\% in channel 4, using the MRS PSF models. We corrected the central spectrum by these flux percentages using the nuclear spectrum. We note that the central spectrum includes the emission of the nuclear and circumnuclear regions, but it is not the sum of both of them.\footnote{The nuclear, circumnuclear, and central NGC~6552 spectra are public available in this \href{https://cloud.cab.inta-csic.es/index.php/s/Ew7nAxNKZEXcEys}{link}}

Figures \ref{spec1.plot} and \ref{spec2.plot} show the nuclear, circumnuclear, and central MRS spectra of NGC~6552, and the identification of all emission lines and polycyclic aromatic hydrocarbon (PAH) features. The nuclear and central spectra are dominated by the central AGN emission, which is mainly composed by a steeply rising continuum due to warm dust emission and a large number of high-excitation and coronal emission lines. The spectra also show low-excitation emission lines, warm hydrogen molecular lines, and PAH features from the interplay between the star-formation and the AGN components of NGC~6552. The presence of PAHs indicate that the AGN radiation field has not depleted or destroyed them like seen in other high-luminosity Seyferts \citep[e.g.][]{Alonso-Herrero+20, Garcia-Bernete+22}. The spectra of the circumnuclear region is mainly dominated by the star-formation component of the galaxy showing low-excitation emission lines, hydrogen molecular lines, and PAH features. However, we found that [NeVI]7.652$\mu$m, [NeV]14.32$\mu$m, and [NeV]24.32$\mu$m coronal emission lines are still present in channels 2, 3 and 4 of the circumnuclear spectrum. The percentage of flux in the circumnuclear spectrum with respect to the nuclear spectrum is 1\%, 6\%, and 11\% (see Table~\ref{table.fluxes}), respectively. Therefore, coronal emission lines and some contribution from the continuum coming from the nuclear region of could still be contaminating the circumnuclear spectrum.

The MRS central spectrum and the \textit{Spitzer} IRS-LL low-resolution spectrum \citep{Jarrett11} of NGC~6552 agree within $\sim$5\% in channels 3 and 4 (see Figure \ref{spec2.plot}). Similar results are found in the MIRI commissioning observation of the planetary nebula SMP-LMC-058 for the full MRS wavelength coverage \citep{Jones+23}. The MIRI Imager F560W photometry of the central region, derived from the unsaturated part of the ramps and in a radius of 1.6 arcsec, is 28$\pm$3 mJy (see Section \ref{MIRI.imager}). The equivalent flux for the MRS, taking the central spectra and performing the convolution with the F560W filter transmission curve, is equal to 24$\pm$3 mJy. Then, the photometry of the central region agree within 10\% for the MIRI Imager F560W image and the MRS channel 1SHORT spectrum. The NGC~6552 calibrated data agree with the statement that the MRS absolute flux uncertainty, based on commissioning data, is $\lesssim$10\% \citep{Rigby22}. Then, we assume a 10\% absolute flux errors in deriving line ratios and analysis that follows.

\section{Results and discussion}\label{disc.sec}

\subsection{Emission line fluxes and kinematics}

NGC~6552 shows a considerable number of emission lines with high signal-to-noise ratios (S/N) in the nuclear spectrum. We detect all Neon lines present in the mid-IR wavelength range, from [NII] to [NeVI], with S/N between 12 to 80. The presence of coronal atomic lines is evident, going from the brightest ones, with S/N>30, [NeVI]7.65$\mu$m and [NeV]14.32$\mu$m, to shallower ones, with S/N between 20 to 50; that is: [MgVII]5.50$\mu$m, [MgV]5.61$\mu$m, [FeVIII]5.45$\mu$m, or [FeVII]7.82$\mu$m. The same happens with high-excitation atomic lines, that is: [ArIII]8.99$\mu$m, [SIV]10.51$\mu$m, [OIV]25.89$\mu$m, and low-excitation atomic lines, i.e. [FeII]5.34$\mu$m, [ArII]6.99$\mu$m, [NeII]12.81$\mu$m, [NeIII]15.56$\mu$m and [SIII]18.71$\mu$m, all of them detected with S/N between 30 to 100. The rotational molecular hydrogen lines from the \Hmol(0-0)S(1) to \Hmol(0-0)S(8) transitions are detected with S/N between 20 to 50, except for \Hmol(0-0)S(7) line that is blended with [MgVII]5.50$\mu$m and its S/N drops to 10. We also detected additional emission lines with lower S/N, the most relevant one being Pfund-$\alpha,$ with S/N of $\sim$14.

We identified and analyzed all detected emission lines with S/N higher than 3 in the NGC~6552 1D extracted spectra. Depending on the line profiles, we performed one-component and two-component Gaussian fit, plus a second-order polynomial to fit the continuum and emission line \citep{Markwardt2009}\footnote{We used the MPFIT Python routine to perform the fits. The version used is available here:  \url{https://github.com/segasai/astrolibpy/tree/master/mpfit}}. The MRS resolving power ranges from 4000 in channel 1 to 1500 in channel 4, corresponding to the line FWHM from 75km\,s$^{-1}$ to 200km\,s$^{-1}$ \citep{Labiano21,Jones+23}. The instrumental line broadening was included in the line profile fitting algorithm. The uncertainties on the derived emission line parameters, like the line FWHM, flux, central wavelength, etc, were estimated using a Monte Carlo simulation. The noise of the spectrum was measured as the root mean square (rms) of the continuum surrounding the emission line. This noise was used to generate new spectra (n=500), where a random Gaussian noise with a sigma equal to the rms was added to the original spectrum before the lines were fitted again. The final uncertainty is the standard deviation of the n individual measurements. Table \ref{table.fluxes} presents the fluxes of all emission lines detected in the nuclear, circumnuclear, and central spectra of NGC~6552. The central wavelengths of the systemic components of all emission lines detected in the nuclear spectra give a median redshift of 0.0267$\pm$0.0002, in agreement with previous estimates \citep{2002LEDA}. We found that the current MRS wavelength calibration is, in general, better than the FWHM of the MRS line spread function (LSF, \citealt{Labiano21,Jones+23}).   

\begin{table}[h]
\caption{Fluxes of all detected emission lines in the nuclear, circumnuclear, and central spectra of NGC~6552.}\label{table.fluxes}
\begin{center}
\resizebox{\hsize}{!}{
\begin{tabular}{cc|ccc}
\hline
  \multirow{2}{*}{Line} & $\lambda_{lab}$ & F$_{Nuclear}$ & F$_{Circum.}$ & F$_{central}$ \\
   & $[\mu$m$]$ & \multicolumn{3}{c}{$[\times 10^{-15}$ erg/s/cm$^2]$} \\
    (1) & (2) & (3) & (4) & (5) \\
\hline
%  \Hmol(1-1)S(9)$^{(1)}$ & 4.954 & 0.17$\pm$0.03 &   & 0.9$\pm$0.2\\
  \Hmol(0-0)S(8)$^{(a)}$ & 5.053 & 2.33$\pm$0.08 & 2.1$\pm$0.7 & 4.2$\pm$0.07\\
  $[$FeII$]^{(b)}$ & 5.340 & 10.4$\pm$0.3 & 11$\pm$2 & 20.9$\pm$0.6\\
  $[$FeVIII$]^{(b)}$ & 5.447 & 7.9$\pm$0.4 &  & 11$\pm$1\\
  $[$MgVII$]^{(b)}$ & 5.503 & 12.4$\pm$0.3 &  & 13$\pm$2\\
  \Hmol(0-0)S(7)$^{(a)}$ & 5.511 & 7.7$\pm$0.6 & 7.0$\pm$0.4 & 12$\pm$2 \\
  $[$MgV$]^{(b)}$ & 5.610 & 16.2$\pm$0.4 &   & 20.1$\pm$0.6\\
%  $[$FeII$]^{(1)}$ & 5.674 & 0.5$\pm$0.1 &    & 0.50$\pm$0.04\\
  $[$KIV$]^{(a)}$ & 5.982 & 1.0$\pm$0.2 &   & 1.0$\pm$0.1\\
  \Hmol(0-0)S(6)$^{(a)}$ & 6.109 & 4.0$\pm$0.2 & 1.6$\pm$0.4 & 6.5$\pm$0.2\\
  $[$SiVII$]^{(a)}$ & 6.492 & 0.9$\pm$0.2 &    & 0.7$\pm$0.2\\
  $[$NiII$]^{(a)}$ & 6.636 & 1.4$\pm$0.3 &    &    \\
  \Hmol(0-0)S(5)$^{(a)}$ & 6.910 & 16.6$\pm$0.5 & 6.6$\pm$0.2 & 29.4$\pm$0.4\\
  $[$ArII$]^{(b)}$ & 6.985 & 43.0$\pm$0.9 & 5.7$\pm$0.3 & 59$\pm$1\\
  $[$NaIII$]^{(a)}$ & 7.318 & 3.1$\pm$0.2 &    & 3.7$\pm$0.3\\
  Pfund-$\alpha^{(a)}$ & 7.460 & 2.7$\pm$0.2 &  & 3.9$\pm$0.1\\
  $[$NeVI$]^{(b)}$ & 7.652 & 70$\pm$2 & 0.6$\pm$0.07 & 77$\pm$2\\
  $[$FeVII$]^{(b)}$ & 7.815 & 5.9$\pm$0.8 &    & 6.9$\pm$0.8\\
  $[$ArV$]^{(b)}$ & 7.902 & 3.3$\pm$0.2 &  & 4.3$\pm$0.4\\
  \Hmol(0-0)S(4)$^{(a)}$ & 8.025 & 8.3$\pm$0.2 & 2.1$\pm$0.1 & 14.2$\pm$0.2\\
  $[$ArIII$]^{(b)}$ & 8.991 & 22.2$\pm$0.5 & 1.2$\pm$0.2 & 27.6$\pm$0.6 \\
  $[$FeVII$]^{(b)}$ & 9.527 & 7.8$\pm$0.4 &  & 8.1$\pm$0.5\\
  \Hmol(0-0)S(3)$^{(a)}$ & 9.66 & 14.3$\pm$0.3 & 5.7$\pm$0.2 & 27.0$\pm$0.4\\
  $[$SIV$]^{(b)}$ & 10.51 & 52$\pm$1 & 6.1$\pm$0.4 & 65$\pm$2\\
  \Hmol(0-0)S(2)$^{(a)}$ & 12.28 & 11.0$\pm$0.4 & 4.1$\pm$0.1 & 16.5$\pm$0.3\\
  $[$NeII$]^{(b)}$ & 12.81 & 143$\pm$3 & 11.3$\pm$0.3 & 160$\pm$4\\
  $[$NeV$]^{(b)}$ & 14.32 & 59$\pm$2 & 3.5$\pm$0.2 & 63$\pm$2\\
  $[$NeIII$]^{(b)}$ & 15.56 & 158$\pm$2 & 11.3$\pm$0.2 & 169$\pm$3\\
  \Hmol(0-0)S(1)$^{(a)}$ & 17.03 & 22.6$\pm$0.6 & 8.6$\pm$0.2 & 31.3$\pm$0.2\\
  $[$SIII$]^{(b)}$ & 18.71 & 67$\pm$1 & 7.2$\pm$0.3 & 73$\pm$2\\
  $[$NeV$]^{(b)}$ & 24.32 & 34$\pm$3 & 3.7$\pm$0.5 & 35$\pm$2\\
  $[$OIV$]^{(b)}$ & 25.89 & 149$\pm$5 & 18$\pm$1 & 155$\pm$5\\
\hline
\end{tabular}}
\end{center}
\tablefoot{Columns describe the; (1) name of each detected emission line, (2) central wavelength from lab of each emission line, and (3,4,5) measured flux in the nuclear, circumnuclear, and central spectra, respectively. $^{(a)}$ Line fit with one Gaussian component. $^{(b)}$ Line fit with two Gaussian components. Nuclear region covers a circular area centred on the nucleus of the source and with a radius equal to $1.5 \times FWHM (\lambda)$, where $FWHM (\lambda) = 0.17$ kpc for $\lambda < 8\mu$m and $FWHM (\lambda) = 0.17\times \lambda[\mu m] / 8$ kpc for $\lambda > 8\mu$m. Circumnuclear region covers an annulus area centred on the nucleus of the source, and with an inner radius of 0.55\,kpc and an outer radius of 0.88\,kpc. Central region covers a circular area centred on the nucleus of the source and with a radius of 0.88\,kpc.} Additional uncertainties in the absolute flux of 10\% for all channels should be considered.
\end{table}

\begin{table*}[h]
\caption{Properties of the blue-shifted velocity component for all high-excitation and coronal emission lines in the nuclear spectrum of NGC~6552.}\label{table.coronal}
\begin{center}
\begin{tabular}{lccccccc}
\hline
 Line & $\lambda_{lab}$ & E$_{ion}^{*}$ & $\log(n_{crit})$& Flux & FWHM & Vpeak & Vout \\ 
    & $[\mu$m$]$ &  [eV] & cm$^{-3}$ & [\%] & [km/s] & [km/s] & [km/s] \\ 
 (1) & (2) & (3) & (4) & (5) & (6) & (7) & (8) \\
\hline
  $[$FeVIII$]$& 5.447 & 125.0 & 6.41  & 0.61 $\pm$ 0.04 & 637 $\pm$ 67 & -88 $\pm$ 19 & 629 $\pm$ 61 \\
  $[$MgVII$]$ & 5.503 & 186.5 &  6.53 & 0.77 $\pm$ 0.02 & 786 $\pm$ 37 & -36 $\pm$ 21 & 704 $\pm$ 25 \\
  $[$MgV$]$ & 5.610 & 109.2 & 6.60  & 0.67 $\pm$ 0.04 & 661 $\pm$ 22 & -167 $\pm$ 12 & 728 $\pm$ 22 \\
  $[$NeVI$]$ & 7.652 & 126.2 & 5.80 & 0.63 $\pm$ 0.03 & 579 $\pm$ 20 & -151 $\pm$ 10 & 642 $\pm$ 22 \\
%  $[$FeVII$]$ & 7.815 & 99.1 & 6.1 & 0.66 $\pm$ 0.06 & 642 $\pm$ 71 & -154 $\pm$ 49 & 700 $\pm$ 57 \\
  $[$ArV$]$ & 7.902 & 59.8 & 5.20 & 0.71 $\pm$ 0.07 & 579 $\pm$ 61 & -175 $\pm$ 21 & 666 $\pm$ 51 \\
  $[$ArIII$]$ & 8.991 & 27.6 & 5.28 & 0.67 $\pm$ 0.03 &  683 $\pm$ 21 & -151 $\pm$ 11 & 731 $\pm$ 19 \\
  $[$FeVII$]$ & 9.527 & 99.1 & 5.74 & 0.75 $\pm$ 0.02 & 980 $\pm$ 35 & -79 $\pm$ 17 & 911 $\pm$ 26 \\
  $[$SIV$]$ & 10.51 & 34.8 & 4.75 & 0.70 $\pm$ 0.01 & 742 $\pm$ 20 & -98 $\pm$ 7 & 727 $\pm$ 17 \\
  $[$NeV$]$ & 14.32 & 97.1 & 4.51 & 0.66 $\pm$ 0.04 & 606 $\pm$ 21 & -175 $\pm$ 13 & 689 $\pm$ 23 \\
  $[$NeIII$]$ & 15.56 & 41.0 & 5.32 & 0.57 $\pm$ 0.02 & 652 $\pm$ 23 & -117 $\pm$ 10 & 672 $\pm$ 19 \\
%  $[$NeV$]$24.32$\mu$m & 97.1 & 3.77 & 0.67 $\pm$ 0.03 & 629 $\pm$ 19 & -180 $\pm$ 12 & 715 $\pm$ 22 \\
  $[$OIV$]$ & 25.89  & 54.9 & 4.00 & 0.59 $\pm$ 0.13 & 493 $\pm$ 102 & -165 $\pm$ 75 & 584 $\pm$ 25 \\
\hline\end{tabular}
\end{center}
\tablefoot{Columns describe the; (1) name of each high-excitation and coronal emission line, (2) intrinsic wavelength from lab, (3) ionisation potential, (4) critical density, (5) percentage of flux in the blue-shifted component, (6) FWHM of the blue-shifted component, (7) velocity offset relative to the narrow component, and (8) the outflow maximal velocity. [NeV]24.32$\mu$m line profile is affected by noise and fringes, and $[$FeVII$]$7.815$\mu$m by a spike, both of them have been discarded from the characterisation of the blue-shifted velocity component.}
\end{table*}

All the emission lines in the NGC~6552 spectra are spectrally resolved. Molecular hydrogen lines present one-component Gaussian profiles, with intrinsic line FWHMs of 312$\pm$34km\,s$^{-1}$. In contrast, the high S/N atomic forbidden lines (high- or low-excitation ones) have asymmetric profiles, with blue wings. Pfund-$\alpha$, the only hydrogen atomic line detected in the NGC~6552 spectra, presents an intrinsic line FWHM of 442$\pm$22km\,s$^{-1}$, greater than the average widths of the atomic forbidden and molecular hydrogen lines, and with no evidence of asymmetries.  

\begin{figure*}[t]
\begin{center}
 \includegraphics[width=\hsize]{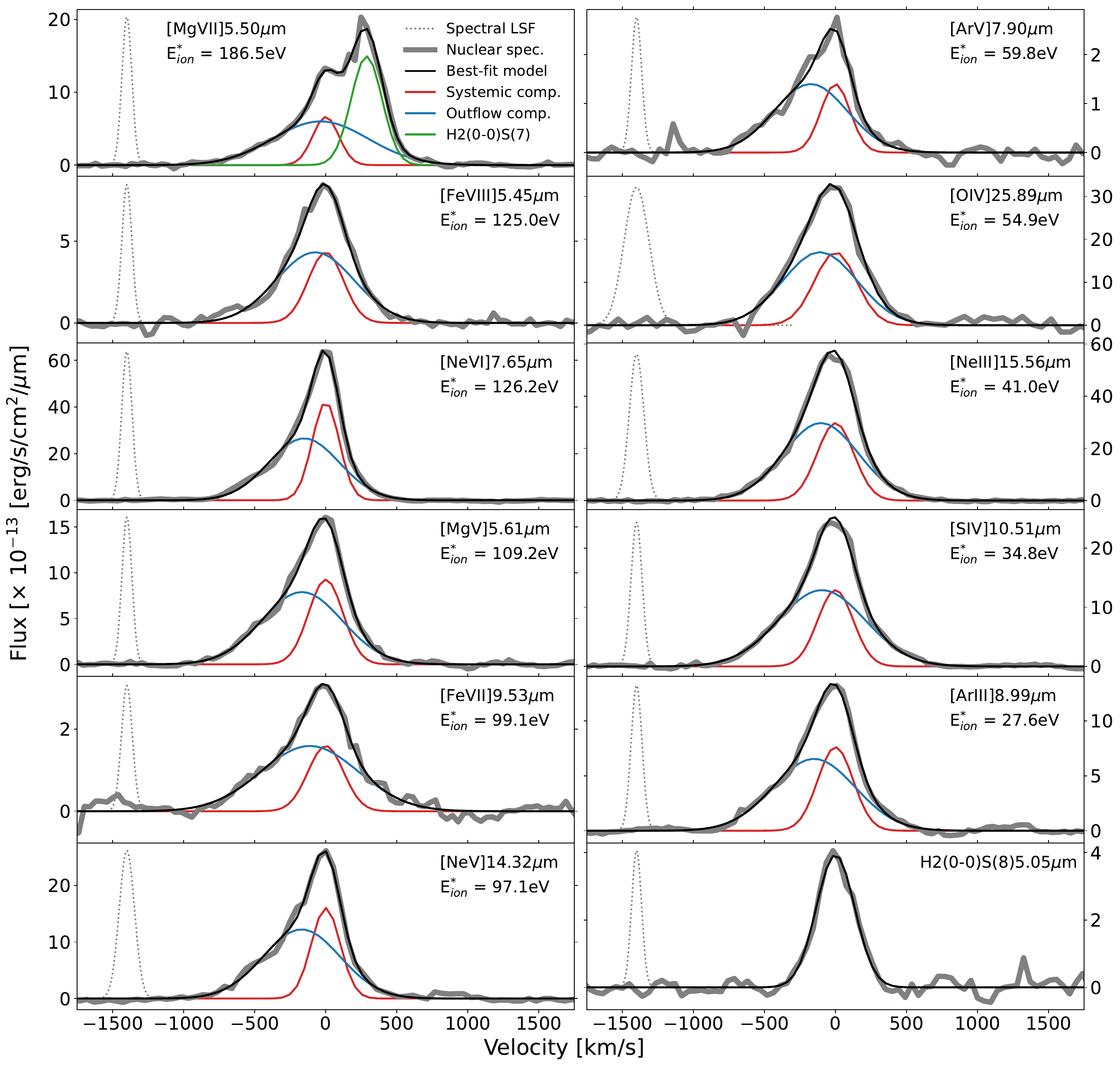} 
\caption{Characterisation of line profiles of high-excitation and coronal emission lines in the nuclear spectrum of NGC~6552. The two-component Gaussian fit of eleven coronal emission lines covering ionisation potential from 27.6 to 186.5 eV is shown. If more than two emission lines trace the same ion (e.g. [NeV] at 14 and 24$\mu$m), only the line with the highest spectral resolution is shown. For comparison, the bottom-right panel illustrates the one-component Gaussian fit of the \Hmol(0-0)S(8) hydrogen molecular line. Gray line shows the nuclear spectrum of NGC~6552. Black line is the best line-fit model. Red line is the systemic component. Blue line is the outflow component. Dotted gray line is the unresolved emission line for the corresponding wavelength. All systemic and outflow components of high-excitation and coronal emission lines are consistent independently of the ionisation potential. The molecular hydrogen lines do not present an outflow component, and its line profiles are in agreement with the systemic component of all high-excitation and coronal emission lines.}\label{coronal_line_fit} 

\end{center}

\end{figure*}

The atomic lines show a combination of systemic and a blue-shifted velocity components and we characterised them by a two-component Gaussian fit. Figure \ref{coronal_line_fit} illustrates the analyses of high-excitation and coronal atomic lines (E$_{ion}^{*} \gtrsim$ 30 eV). We found that the systemic and blue-shifted velocity components of these emission lines are consistent within the given error bars, independent of their ionisation potentials or critical densities. The average systemic line FWHM, 270$\pm$50 km\,s$^{-1}$, agrees with the one of the molecular hydrogen lines. The blue-shifted velocity components are not spatially resolved in the MRS observations, suggesting that they are fully localised in the nucleus of the galaxy. Its physicals scales should be lower than 170pc that corresponds to the physical scales of the MRS PSF FWHM at 5$\mu$m.  Table \ref{table.coronal} summarises the properties of the blue-shifted velocity component, together with the ionisation potential and critical density of each high-excitation atomic line in the nuclear spectrum of NGC~6552. The blue-shifted velocity components is characterised by average velocity offsets (V$_{peak}$) of -127$\pm$45 km\,s$^{-1}$ relative to the narrow component and with intrinsic line FWHMs of 673$\pm$123 km\,s$^{-1}$. The nuclear blue-shifted velocity component dominates the emission of the high-excitation lines with a 67$\pm$7\% of the total flux of the line. Low-excitation atomic lines (E$_{ion}^{*} \lesssim$ 20 eV, namely, [FeII], [ArII], [SIII], and [NeII]) also show the presence of a blue-shifted velocity component with similar kinematic properties, that is, V$_{peak}$ and line FWHMs, as those identified in the high-excitation lines but carrying out only between 20\% to 36\% of the total flux of the line. Then, their main flux contribution is associated with the systemic velocity component that shows a slightly broader line FWHMs (354$\pm$26 km\,s$^{-1}$) than the high-excitation lines.

\subsection{Highly ionised nuclear outflow. Kinematics and physical properties.}

The blue-shifted velocity components identified in all atomic emission lines are interpreted as being due to the presence of outflowing material close to the AGN. This is the first clear observational evidence for a nuclear outflow in NGC~6552. The outflow share the same kinematics and fraction of the total flux in all high-excitation and coronal emission. This clearly indicates that independently of the ionisation state (from about 28 to 187 eV, see Table \ref{table.coronal}), the lines are tracing the same regions of the nuclear outflow. This is fully ionised, with no evidence of any stratification in its ionisation structure. Moreover, the outflowing gas is characterised by a blue-shifted velocity offset of -127$\pm$45 km\,s$^{-1}$, on average, with outflow maximal velocities (V$_{peak}$ + 2$\times \sigma_{blue}$, where $\sigma_{blue}$ is the sigma of the Gaussian component fit) of 698$\pm$80 km\,s$^{-1}$ (see Table \ref{table.coronal}). This implies that when assuming a bi-conical structure centred on the AGN, we are predominantly seeing the outflowing gas coming towards us. The opacity of the interstellar medium in the mid-IR is significantly smaller (factors 30 to 100) than that in the optical. However, for the large column densities measured towards the AGN in NGC6552,
i.e. $\log(N_{H})$ = 24.05 cm$^{-3}$ \citep{Ricci2015}, a large optical depth between 3.3 and 10 is expected in the
5 to 25 $\mu$m MRS spectral range (e.g. \citealt{Corrales+16}). Therefore the lack of a red velocity component in
the emission line profile is interpreted as emitting outflowing gas moving away from us (i.e. receding direction)
but obscured from our line of sight, even at these wavelengths, by the torus around the AGN and/or the dense
medium in its vicinity.

The ratio of the [NeV]14.32$\mu$m to [NeV]24.32$\mu$m emission lines  gives an estimation of the electron density in the highly ionised coronal emission gas. Independently of the unknown electron temperature, the low ratio measured in NGC 6552 (1.7$\pm$0.4, using absolute flux errors), indicates that the coronal gas is in the low density regime, that is, less than 10$^{5.5}$ cm$^{-3}$ for temperatures less than 10${^6}$ K \citep{Dudik+07}. However, the kinematics of the high-excitation and coronal gas (27.6 < E$_{ion}^{*} [eV] <$ 186.5) are very similar, while the critical electron densities cover a wide range from $\log(n_{crit}[cm^{-3}])$ equal to 4 to 6.6 (see Figure \ref{coronal_line_fit} and Table \ref{table.coronal}). This indicates that the physical conditions of the ionised gas in the outflow must be closer to the lower densities  (i.e. less than a few 10$^3$ cm$^{-3}$) and temperatures (i.e. less than few 10$^4$ K). Otherwise, lines such as [NeV]14.3$\mu$m and [OIV]25.9$\mu$m would not be detected as strong as presented in the NGC~6552 nuclear spectrum \citep{Pereira-Santaella+10}. The lack of [FeII]4.89$\mu$m emission is also consistent with low denisities \citep[$<10^3$cm$^ {-3}$][]{Pereira22}.

The outflowing material appears only in the ionised gas, with no evidence of the outflow in any of the molecular rotational lines present in the nuclear spectrum. In fact, the average line FWHMs (312$\pm$34 km\,s$^{-1}$) of the molecular lines agree with the systemic velocity component of the ionised gas (270 $\pm$ 50 km s$^{-1}$). This could indicate that: (i) the molecular gas in the nuclear region is not directly facing the radiation coming out of the AGN and its velocity field is determined by the dynamical mass in the nuclear region or (ii) when the molecular hydrogen becomes part of the outflow, it gets quickly dissociated, contributing to wider the Pfund-$\alpha$ hydrogen line.

\subsection{Warm molecular hydrogen}
\label{WMH}

The JWST and, in particular, the MRS, will revolutionise our ability to observe the infrared lines of molecular hydrogen \citep{Guillard+15}. In NGC~6552, we  detected a suite of pure rotational lines, from the brightest 0--0~S(1) 17~\mum\ transition up to the 0--0~S(8) 5.05~\mum\ line, at remarkable high S/N for each of the three spectral extractions (central, nuclear, and circumnuclear). Those lines arise from warm ($\approx 100 - 20000$~K) molecular gas \citep{Habart+05}. We have used these observed line fluxes (Table~\ref{table.fluxes}) to derive the column densities and masses of warm \Hmol\ for each of the three regions by fitting the \Hmol\ excitation diagrams \citep[see][for a review]{Wakelam+17}. The statistical errors from the fits, listed in Table~\ref{table.fluxes}, have been added in quadrature with a 10\% absolute flux uncertainty.

We emphasise that the warm ($\gtrsim 100$~K) gas is only a fraction of the total molecular mass. Most of the cold molecular gas cannot be traced by the mid-infrared \Hmol\ lines. If the 0--0~S(0) line at $\lambda = 28.2$~\mum, which is located at the far end of the MRS-covered wavelength range and currently not properly calibrated, were available, it would provide access to the lower temperature gas and a better constraint on the total warm \Hmol\ mass. Hence, the numbers derived here can be viewed as lower limits \citep[see][for a discussion]{Guillard+12}. 

\begin{figure}[t]
\begin{center}
\includegraphics[width=\hsize]{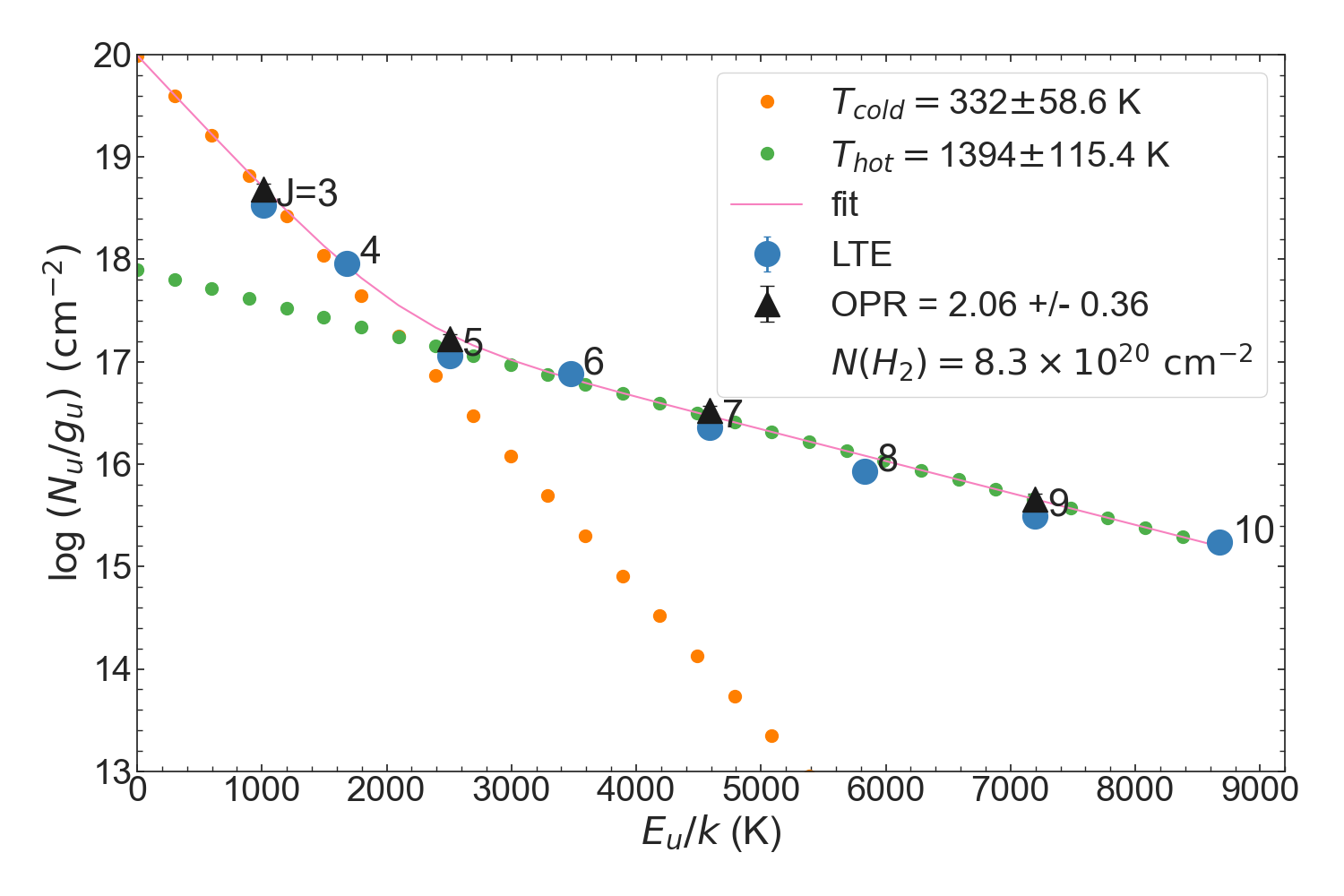}
\includegraphics[width=\hsize]{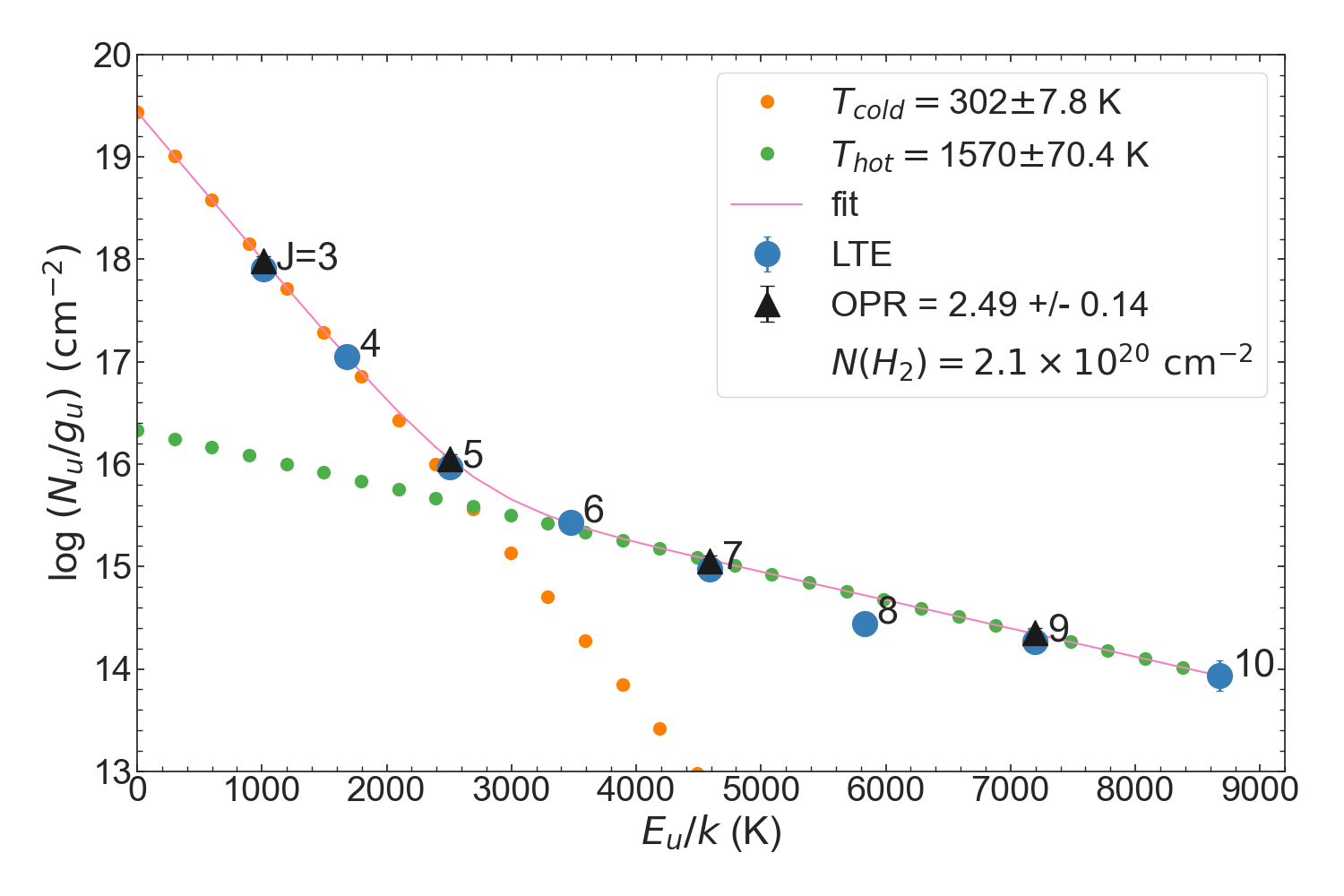}
\includegraphics[width=\hsize]{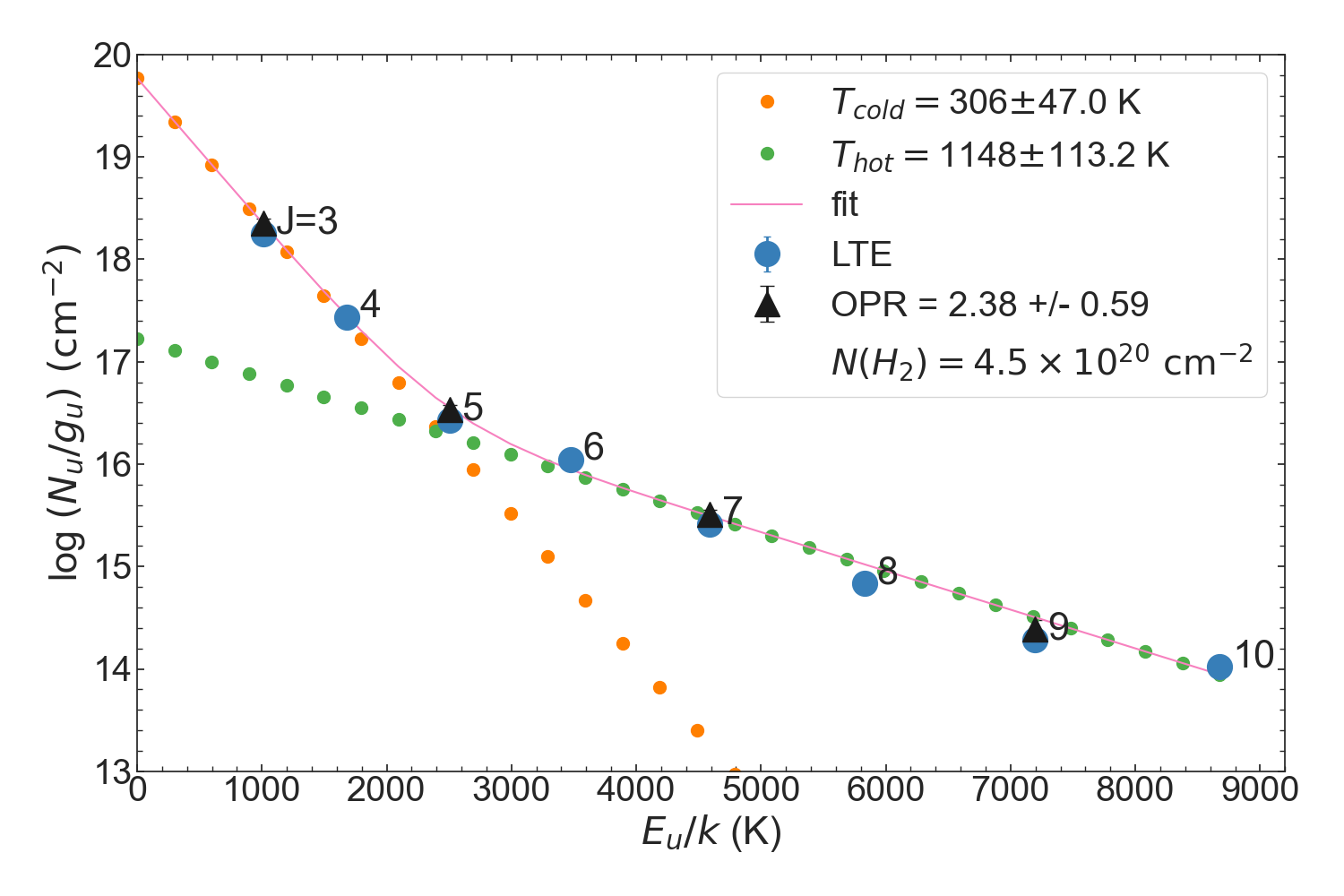}
\caption{NGC6552 \Hmol\ excitation diagrams of the three extracted regions (\textit{top}: nucleus, \textit{middle}: circumnuclear, and \textit{bottom}: central). The logarithm of the column densities divided by the statistical weights of the different transitions are plotted against the upper level energies of those transitions. The LTE and fitted-OPR column densities are both displayed: the blue circles are the original column densities with LTE $g_u$, assuming OPR=3, while the black triangles represent the column densities calculated using the fitted OPR (see Sect. \ref{WMH} for details).  The \Hmol\ temperatures, column densities, and masses are listed in Table~\ref{table:H2param}.}
\label{fig:H2}
\end{center}
\end{figure}

Figure~\ref{fig:H2} shows the excitation diagrams for the three spectral extractions, where the logarithm of the column densities of the upper \Hmol\ levels divided by their statistical weights, $ln(N_u /g_u )$, are plotted against their excitation energies, $E_u/k_B$, expressed in K. For a single uniform temperature $ln(N_u /g_u ) \propto T^{-1}_{exc}$, where $T_{exc}$ is the excitation temperature. The plotted values of $ln(N_u /g_u )$ have been constructed in two ways. First, in a situation of local thermal equilibrium (LTE), where the excitation temperature equals that of the gas, and where the ortho-to-para ratio ($OPR$) is assumed to be $OPR=3$. Those values, shown as the blue circles on Figure~\ref{fig:H2}, exhibit a classical curvature, which indicates that multiple temperatures are present, and we get good agreement with a two-temperature fit that invokes a warm component (typically $\approx 300$~K here) and a hot component ($\approx 1200$~K). We note that the spectral extraction in the nucleus shows the highest excitation. It is expected because of the stronger radiation field, and possibly stronger cosmic ray ionisation rate, or enhanced dissipation of turbulent energy close to the central nuclei \citep{Ogle+10}. The distribution of the $ln(N_u /g_u )$ values also exhibits a 'zigzag' pattern which shows that the ortho-to-para ratio is smaller than 3 for all the regions. This indicates that the \Hmol\ gas is thermalised to lower temperatures \citep{Flagey+13}, or that its  chemistry is out of equilibrium, because the time spent hot for the gas is shorter than the ortho-to-para conversion time, as is the case in molecular shocks for instance \citep[e.g.][]{Neufeld+98, Wilgenbus+00}. We therefore constructed the excitation diagram in a second way, by fitting the OPR in addition to the two temperatures. Those column densities are displayed as the black triangles and show a smoother fit to the data. We used those fits to estimate the physical parameters of the warm \Hmol\ gas. 

The legends of each panel in Fig.~\ref{fig:H2} shows the results of the temperature and OPR fits, as well as the total column densities. Additionally, all the physical parameters derived from the fits of the excitation diagrams are gathered in Table~\ref{table:H2param}. The warm \Hmol\ masses are derived assuming an angular distance of 120~Mpc and solid angles corresponding to the extracted regions as stated in Sect.~\ref{Spectra.sec}. To summarise, a mass of warm \Hmol\ of at least $1.9 \times 10^7~\Msun$ is present in the central region (1.8~kpc in diameter), with almost 30\% of that mass in the circum-nuclear region.  The masses quoted in Table~\ref{table:H2param} are typical of what is detected at these temperatures in nearby radio galaxies or infrared galaxies \citep[e.g.][]{Guillard+12, Petric+18}.
Again, if we had access to the 0--0~S(0) line, which is sensitive to cooler gas, or if we had performed a more sophisticated modelling taking into account a distribution of temperatures, as in \citet{Togi+16} for instance, our derived \Hmol\ masses would be a factor 2-10 larger. We defer to a future paper the line maps fitting with detailed physical models. 

\begin{table}[h]
\caption{Physical parameters derived from the two-temperatures fitting of the \Hmol\ excitation diagrams for the three spectral extractions. The error bars on those parameters are calculated by propagating the uncertainties on the line fluxes, which include a 10\% absolute flux calibration error and fitting errors added in quadrature.}
\label{table:H2param}
\begin{center}
\resizebox{\hsize}{!}{
\begin{tabular}{lcccc}
\hline
Parameter & & Nuclear & Circum. & Central \\
\hline
T(cold) [K]                    & (1) & $332 \pm 59$   & $302 \pm 8$     & $306 \pm 47$ \\
T(hot)  [K]                    & (2) & $1394 \pm 115$ & $1570 \pm 70$   & $1148 \pm 113$ \\
N(cold) [10$^{20}$ cm$^{-2}$]  & (3) & $8.0 \pm 5.8 $ & $2.05 \pm 0.29$ & $4.4 \pm 3.4$ \\
N(hot)  [10$^{19}$ cm$^{-2}$]  & (4) & $2.8 \pm 1.1 $ & $0.08 \pm 0.01$ & $0.48 \pm 0.27 $ \\
N(total) [10$^{20}$ cm$^{-2}$] & (5) & $8.3 \pm 5.8 $ & $2.1 \pm 0.3$   & $4.5 \pm 3.4 $\\
Ortho-to-para ratio            & (6) & $2.06 \pm 0.36$& $2.5 \pm 0.2$   & $2.38\pm 0.59$ \\
M(\Hmol) [10$^7$ \Msun]        & (7) & $1.38 \pm 0.79$& $0.56 \pm 0.12$ & $1.95 \pm 1.13$ \\
\hline\end{tabular}}
\end{center}
\tablefoot{Rows describe each parameter for the nuclear, circumnuclear, and central regions. The parameters are the; (1) temperature of the cold component, (2) temperature of the hot component, (3) column density associated with the cold component, (4) column density associated with the hot component, (5) column density associated with the sum of the cold and hot component, (6) ortho-to-para ratio, and (7) total \Hmol\ mass.}
\end{table}

\subsection{Seyfert nature of the AGN and black hole mass}\label{discuss.BH}

The mid-IR emission lines of ionised gas have been used to disentangle the AGN and stellar emission in nearby galaxies, in particular, the sequence of different ionisation levels of neon ([NeII]12.82$\mu$m, [NeIII]15.56$\mu$m to [NeV]14.32,24.32$\mu$m), and the [OIV]25.89$\mu$m line \citep{Pereira-Santaella+10}. Table \ref{table.ratios} shows the Neon and Oxygen line ratios for NGC~6552. They confirm that the nuclear, circumnuclear, and central spectra of NGC~6552 are close to the median values of Seyfert 1 and 2, indicating that the ionisation of the interstellar medium in NGC~6552 at distances up to a radius of 0.88kpc from the nucleus is dominated by the AGN radiation field. However, in Section \ref{Spectra.sec}, we present our conclusion that channels 3 and 4 spectra of the circumnuclear region could contain up to 6\% and 11\% of the nuclear AGN emission,  making it difficult to disentangle if the circumnuclear region is dominated by the radiation field of the AGN or star formation.  

The black hole (BH) mass based on the line FWHM of the high-excitation emission line was obtained using \textit{Spitzer} high-resolution spectroscopy in nearby galaxies (e.g. \citealt{Dasyra+08}). Taking the line FWHM of the systemic component in the nuclear spectrum of [NeV]14.3$\mu$m (205$\pm$15 km\,s$^{-1}$), we derived a NGC~6552 BH mass, $\log(M_{BH}[M_{\sun}]$), of 6.3$\pm$0.5. This gives a mass range of 0.6 to 6 million solar masses, which is (on average) the mass of the Milky Way BH (4.1$\times$10$^{6}$M$_{\sun}$, \citealt{Ghez2008, EHTC_L12+22}). As the MRS has a spectral resolution five times higher than the \textit{Spitzer} high resolution mode and a sub-arcsec angular resolution at all wavelengths, the line profiles of the high-excitation and coronal lines with MIRI will provide far better and more accurate measurements of the BH mass than the initial estimates based on previous mid-IR spectroscopy. 

The only hydrogen recombination line detected in the mid-IR spectral of NGC~6552 is Pfund-$\alpha$ with a line FWHM of 442 $\pm$ 22 km s$^{-1}$, which is a factor 1.5-2 larger than the ionised and molecular lines, and shows no evidence of outflowing gas. This result can be interpreted in two  ways. On the one hand, the broader Pfund-$\alpha$ line could be tracing the inner broad line region (BLR) in this galaxy. There is already some evidence of a hidden BLR based on polarised light measurement \citep{Tran+01}. However, the S/N of Pfund-$\alpha$ is not good enough to trace a very broad weak line as expected if the BLR is mostly obscured in this CT-AGN. On the other hand, the line FWHM could represent the dynamical mass of the host galaxy contained within the aperture while the width of the molecular and ionised gas would be tracing the dynamical mass in smaller regions, closer to the AGN. 

\begin{table}[h]
\caption{Observed line ratios for NGC~6552 regions}\label{table.ratios}
\begin{center}
\begin{tabular}{lccc}
\hline
Line ratio & Nuclear & Circum. & Central \\
\hline
$[$OIV$]$25.89 / $[$NeV$]$24.32 & 4.4$\pm$1.1 & 4.9$\pm$1.4 & 4.4$\pm$1.0 \\
$[$OIV$]$25.89 / $[$NeV$]$14.32 & 2.5$\pm$0.6 & 5.1$\pm$1.2 & 2.5$\pm$0.6 \\
$[$NeV$]$24.32 / $[$NeV$]$14.32 & 0.6$\pm$0.2 & 1.1$\pm$0.3 & 0.6$\pm$0.2 \\
$[$NeIII$]$15.56 / $[$NeV$]$14.32 & 2.7$\pm$0.6 & 3.2$\pm$0.7 & 2.7$\pm$0.6 \\
$[$NeII$]$12.81 / $[$NeV$]$14.32 & 2.4$\pm$0.5 & 3.2$\pm$0.7 & 2.5$\pm$0.6 \\
$[$NeIII$]$15.56 / $[$OIV$]$25.89 & 1.1$\pm$0.2 & 0.6$\pm$0.2 & 1.1$\pm$0.3 \\
$[$NeII$]$12.81 / $[$OIV$]$25.89 & 1.0$\pm$0.2 & 0.6$\pm$0.2 & 1.0$\pm$0.2 \\
$[$NeIII$]$15.56 / $[$NeII$]$12.81 & 1.1$\pm$0.2 & 1.0$\pm$0.2 & 1.1$\pm$0.2 \\
\hline\end{tabular}
\end{center}
%Note: The uncertainties are calculated using the absolute errors.
\end{table}

\section{Summary and conclusions}\label{sum.con.sec}

Galaxy NGC~6552, with an already identified Compton-thick AGN in its centre \citep{Ricci2015}, was observed with the MIRI Imager and Medium Resolution Spectrometer during the \textit{JWST} commissioning to characterise the persistence of the MIRI detectors. We present the calibrated NGC~6552 MIRI image and MRS spectra, including an extensive and detailed explanation of the MIRI data and calibration process. 

We obtained the nuclear, circumnuclear, and central mid-IR spectra of NGC~6552. The nuclear and central spectra are dominated by the AGN emission, with steeply rising continuum and a large number of high- and low-excitation emission lines, warm molecular Hydrogen lines, and PAH features from the interplay between the star-formation and AGN components. The spectra of the circumnuclear region is mainly dominated by the star-formation, with low excitation lines, molecular hydrogen lines, and PAH features. The central MRS spectra of NGC~6552 is consistent with previous \textit{Spitzer} low-resolution observations and with the MIRI F560W Imager photometry. 

The MIRI IFS provides the first clear observational evidence for a nuclear outflow in NGC~6552. The nuclear AGN is ionising the surrounding regions and producing a blueshifted, high-speed outflow with offset velocities of -127$\pm$45\,km s$^{-1}$ and maximal velocities of 698$\pm$80 km s$^{-1}$. The outflow is not spatially resolved (<0.2kpc) and represents the 67$\pm$7\% of the total emission of high-excitation and coronal emission lines. The analysis suggests a scenario where the outflow is fully ionised, with no evidence of stratification in its ionised structure, and produced in a low-density (<10$^{3}$ cm$^{-3}$) environment. The lack of the red component in the spectra suggests that the receding side of the outflow is obscured by material around the nucleus of the galaxy. Additionally, we confirm that NGC~6552 is a Seyfert galaxy and contains an active black hole with a low-to-intermediate mass ranging from 0.6 to 6 million M$_{\sun}$. 

From two-temperature fits of the \Hmol\ excitation diagrams constructed for the three regions, we derived a warm \Hmol\ mass of at least $1.9 \pm 1.1 \times 10^7~\Msun$ in the central region (1.8~kpc in diameter) of the galaxy, with 30\% of that mass in the circumnuclear region (in an annulus between 0.55 and 0.88~kpc in radius). The \Hmol\ excitation is significantly stronger in the nuclear region. The warm \Hmol\ lines are spectrally resolved, exhibit Gaussian profiles and show no evidence of outflowing gas. The FWHM of the lines is consistent with the systemic component in the high excitation lines, suggesting that the warm molecular gas is somehow shielded from (or not aligned with) the AGN radiation and its kinematics are determined by the BH and stellar dynamical mass. 

This early commissioning paper already demonstrates the huge gain in the scientific performance that JWST/MIRI provides over previous ground- and space-based infrared observatories. With respect to the study of AGNs, the high angular resolution of the MIRI IFS allows us to spatially separate the central regions of active galaxies; the high spectral resolution of MIRI enables kinematic studies that are relatively unaffected by dust extinction. Most importantly, the high sensitivity provided by the 6.5m aperture of the JWST, offers for the first time full access to the zoo of important diagnostic lines, covering a wide range of ionisation states and offering redundant spectral information for cross-checks. Although the data for NGC6552 are non-optimally sampled these advantages are realised and used to provide new constraints on the nature of the nucleus and circumnuclear regions. Altogether, we expect that JWST-MIRI will revolutionise the field of AGN research over the coming years.

\begin{acknowledgements}
%       We thank the anonymous referee for her/his very useful recommendations. 
       
JAM, AL, and LC acknowledge support by grant PIB2021-127718NB-100 by the Spanish Ministry of Science and Innovation/State Agency of Research MCIN/AEI/10.13039/501100011033 and by “ERDF A way of making Europe”. PJK acknowledges financial support from the Science Foundation Ireland / Irish Research Pathway programme under Grant Number 21/PATH-S/9360. IA and DG thank the European Space Agency (ESA) and the Belgian Federal Science Policy Office (BELSPO) for their support in the framework of the PRODEX Programme. The work presented is the effort of the entire MIRI team and the enthusiasm within the MIRI partnership is a significant factor in its success. MIRI draws on the scientific and technical expertise of the following organisations: Ames Research Center, USA; Airbus Defence and Space, UK; CEA-Irfu, Saclay, France; Centre Spatial de Liége, Belgium; Consejo Superior de Investigaciones Científicas, Spain; Carl Zeiss Optronics, Germany; Chalmers University of Technology, Sweden; Danish Space Research Institute, Denmark; Dublin Institute for Advanced Studies, Ireland; European Space Agency, Netherlands; ETCA, Belgium; ETH Zurich, Switzerland; Goddard Space Flight Center, USA; Institute d'Astrophysique Spatiale, France; Instituto Nacional de Técnica Aeroespacial, Spain; Institute for Astronomy, Edinburgh, UK; Jet Propulsion Laboratory, USA; Laboratoire d'Astrophysique de Marseille (LAM), France; Leiden University, Netherlands; Lockheed Advanced Technology Center (USA); NOVA Opt-IR group at Dwingeloo, Netherlands; Northrop Grumman, USA; Max-Planck Institut für Astronomie (MPIA), Heidelberg, Germany; Laboratoire d’Etudes Spatiales et d'Instrumentation en Astrophysique (LESIA), France; Paul Scherrer Institut, Switzerland; Raytheon Vision Systems, USA; RUAG Aerospace, Switzerland; Rutherford Appleton Laboratory (RAL Space), UK; Space Telescope Science Institute, USA; Toegepast- Natuurwetenschappelijk Onderzoek (TNO-TPD), Netherlands; UK Astronomy Technology Centre, UK; University College London, UK; University of Amsterdam, Netherlands; University of Arizona, USA; University of Bern, Switzerland; University of Cardiff, UK; University of Cologne, Germany; University of Ghent; University of Groningen, Netherlands; University of Leicester, UK; University of Leuven, Belgium; University of Stockholm, Sweden; Utah State University, USA. A portion of this work was carried out at the Jet Propulsion Laboratory, California Institute of Technology, under a contract with the National Aeronautics and Space Administration. We would like to thank the following National and International Funding Agencies for their support of the MIRI development: NASA; ESA; Belgian Science Policy Office; Centre Nationale D'Etudes Spatiales (CNES); Danish National Space Centre; Deutsches Zentrum fur Luft-und Raumfahrt (DLR); Enterprise Ireland; Ministerio De Economiá y Competividad; Netherlands Research School for Astronomy (NOVA); Netherlands Organisation for Scientific Research (NWO); Science and Technology Facilities Council; Swiss Space Office; Swedish National Space Board; UK Space Agency. This work is based on observations made with the NASA/ESA/CSA James Webb Space Telescope. The data were obtained from the Mikulski Archive for Space Telescopes at the Space Telescope Science Institute, which is operated by the Association of Universities for Research in Astronomy, Inc., under NASA contract NAS 5-03127 for \textit{JWST}; and from the \href{https://jwst.esac.esa.int/archive/}{European \textit{JWST} archive (e\textit{JWST})} operated by the ESDC. 
        
This research made use of Photutils, an Astropy package for detection and photometry of astronomical sources \citep{larry_bradley_2022_6825092}.

\end{acknowledgements}

\bibliographystyle{aa} % style aa.bst
\bibliography{Bibliography.bib} % your references Yourfile.bib

\begin{thebibliography}{64}
\expandafter\ifx\csname natexlab\endcsname\relax\def\natexlab#1{#1}\fi

\bibitem[{{Alonso-Herrero} {et~al.}(2020){Alonso-Herrero}, {Pereira-Santaella},
  {Rigopoulou}, {Garc{\'\i}a-Bernete}, {Garc{\'\i}a-Burillo},
  {Dom{\'\i}nguez-Fern{\'a}ndez}, {Combes}, {Davies}, {D{\'\i}az-Santos},
  {Esparza-Arredondo}, {Gonz{\'a}lez-Mart{\'\i}n}, {Hern{\'a}n-Caballero},
  {Hicks}, {H{\"o}nig}, {Levenson}, {Ramos Almeida}, {Roche}, \&
  {Rosario}}]{Alonso-Herrero+20}
{Alonso-Herrero}, A., {Pereira-Santaella}, M., {Rigopoulou}, D., {et~al.} 2020,
  \aap, 639, A43

\bibitem[{{{\'A}lvarez-M{\'a}rquez} {et~al.}(2021){{\'A}lvarez-M{\'a}rquez},
  {Marques-Chaves}, {Colina}, \& {P{\'e}rez-Fournon}}]{Alvarez-Marquez+21}
{{\'A}lvarez-M{\'a}rquez}, J., {Marques-Chaves}, R., {Colina}, L., \&
  {P{\'e}rez-Fournon}, I. 2021, \aap, 647, A133

\bibitem[{{Arribas} {et~al.}(2014){Arribas}, {Colina}, {Bellocchi}, {Maiolino},
  \& {Villar-Mart{\'\i}n}}]{Arribas+14}
{Arribas}, S., {Colina}, L., {Bellocchi}, E., {Maiolino}, R., \&
  {Villar-Mart{\'\i}n}, M. 2014, \aap, 568, A14

\bibitem[{Bouchet {et~al.}(2015)Bouchet, Garc{\'{\i}}a-Mar{\'{\i}}n, Lagage,
  Amiaux, Augu{\'{e}}res, Bauwens, Blommaert, Chen, Detre, Dicken, Dubreuil,
  Galdemard, Gastaud, Glasse, Gordon, Gougnaud, Guillard, Justtanont, Krause,
  Leboeuf, Longval, Martin, Mazy, Moreau, Olofsson, Ray, Rees, Renotte,
  Ressler, Ronayette, Salasca, Scheithauer, Sykes, Thelen, Wells, Wright, \&
  Wright}]{Bouchet15}
Bouchet, P., Garc{\'{\i}}a-Mar{\'{\i}}n, M., Lagage, P.-O., {et~al.} 2015,
  PASP, 127, 612

\bibitem[{{Bower} {et~al.}(1996){Bower}, {Hasinger}, {Castander},
  {Aragon-Salamanca}, {Ellis}, {Gioia}, {Henry}, {Burg}, {Huchra}, {Bohringer},
  {Briel}, \& {McLean}}]{Bower96}
{Bower}, R.~G., {Hasinger}, G., {Castander}, F.~J., {et~al.} 1996, \mnras, 281,
  59

\bibitem[{Bradley {et~al.}(2022)Bradley, Sip{\H o}cz, Robitaille, Tollerud,
  Vin{\'{\i}}cius, Deil, Barbary, Wilson, Busko, G{\"u}nther, Cara, Conseil,
  Bostroem, Droettboom, Bray, Bratholm, Lim, Barentsen, Craig, Pascual, Perren,
  Greco, Donath, de~Val-Borro, Kerzendorf, Bach, Weaver, D'Eugenio, Souchereau,
  \& Ferreira}]{larry_bradley_2022_6825092}
Bradley, L., Sip{\H o}cz, B., Robitaille, T., {et~al.} 2022, astropy/photutils:
  1.5.0

\bibitem[{Bushouse {et~al.}(2022)Bushouse, Eisenhamer, Dencheva, Davies,
  Greenfield, Morrison, Hodge, Simon, Grumm, Droettboom, Slavich, Sosey, Pauly,
  Miller, Jedrzejewski, Hack, Davis, Crawford, Law, Gordon, Regan, Cara,
  MacDonald, Bradley, Shanahan, \& Jamieson}]{bushouse_howard_2022_6984366}
Bushouse, H., Eisenhamer, J., Dencheva, N., {et~al.} 2022, spacetelescope/jwst:
  JWST 1.6.2

\bibitem[{{Corrales} {et~al.}(2016{\natexlab{a}}){Corrales}, {Garc{\'\i}a},
  {Wilms}, \& {Baganoff}}]{Corrales2016}
{Corrales}, L.~R., {Garc{\'\i}a}, J., {Wilms}, J., \& {Baganoff}, F.
  2016{\natexlab{a}}, \mnras, 458, 1345

\bibitem[{{Corrales} {et~al.}(2016{\natexlab{b}}){Corrales}, {Garc{\'\i}a},
  {Wilms}, \& {Baganoff}}]{Corrales+16}
{Corrales}, L.~R., {Garc{\'\i}a}, J., {Wilms}, J., \& {Baganoff}, F.
  2016{\natexlab{b}}, \mnras, 458, 1345

\bibitem[{{Dasyra} {et~al.}(2008){Dasyra}, {Ho}, {Armus}, {Ogle}, {Helou},
  {Peterson}, {Lutz}, {Netzer}, \& {Sturm}}]{Dasyra+08}
{Dasyra}, K.~M., {Ho}, L.~C., {Armus}, L., {et~al.} 2008, \apjl, 674, L9

\bibitem[{{Davies} {et~al.}(2020){Davies}, {F{\"o}rster Schreiber}, {Lutz},
  {Genzel}, {Belli}, {Shimizu}, {Contursi}, {Davies}, {Herrera-Camus}, {Lee},
  {Naab}, {Price}, {Renzini}, {Schruba}, {Sternberg}, {Tacconi}, {{\"U}bler},
  {Wisnioski}, \& {Wuyts}}]{Davies+20}
{Davies}, R.~L., {F{\"o}rster Schreiber}, N.~M., {Lutz}, D., {et~al.} 2020,
  \apj, 894, 28

\bibitem[{{Dudik} {et~al.}(2007){Dudik}, {Weingartner}, {Satyapal}, {Fischer},
  {Dudley}, \& {O'Halloran}}]{Dudik+07}
{Dudik}, R.~P., {Weingartner}, J.~C., {Satyapal}, S., {et~al.} 2007, \apj, 664,
  71

\bibitem[{{Emonts} {et~al.}(2017){Emonts}, {Colina}, {Piqueras-L{\'o}pez},
  {Garcia-Burillo}, {Pereira-Santaella}, {Arribas}, {Labiano}, \&
  {Alonso-Herrero}}]{Emonts+17}
{Emonts}, B.~H.~C., {Colina}, L., {Piqueras-L{\'o}pez}, J., {et~al.} 2017,
  \aap, 607, A116

\bibitem[{{Event Horizon Telescope Collaboration} {et~al.}(2022){Event Horizon
  Telescope Collaboration}, {Akiyama}, {Alberdi}, {Alef}, {Algaba}, {Anantua},
  {Asada}, {Azulay}, {Bach}, {Baczko}, {Ball}, {Balokovi{\'c}}, {Barrett},
  {Baub{\"o}ck}, {Benson}, {Bintley}, {Blackburn}, {Blundell}, {Bouman},
  {Bower}, {Boyce}, {Bremer}, {Brinkerink}, {Brissenden}, {Britzen},
  {Broderick}, {Broguiere}, {Bronzwaer}, {Bustamante}, {Byun}, {Carlstrom},
  {Ceccobello}, {Chael}, {Chan}, {Chatterjee}, {Chatterjee}, {Chen}, {Chen},
  {Cheng}, {Cho}, {Christian}, {Conroy}, {Conway}, {Cordes}, {Crawford},
  {Crew}, {Cruz-Osorio}, {Cui}, {Davelaar}, {Laurentis}, {Deane}, {Dempsey},
  {Desvignes}, {Dexter}, {Dhruv}, {Doeleman}, {Dougal}, {Dzib}, {Eatough},
  {Emami}, {Falcke}, {Farah}, {Fish}, {Fomalont}, {Ford}, {Fraga-Encinas},
  {Freeman}, {Friberg}, {Fromm}, {Fuentes}, {Galison}, {Gammie}, {Garc{\'\i}a},
  {Gentaz}, {Georgiev}, {Goddi}, {Gold}, {G{\'o}mez-Ruiz}, {G{\'o}mez}, {Gu},
  {Gurwell}, {Hada}, {Haggard}, {Haworth}, {Hecht}, {Hesper}, {Heumann}, {Ho},
  {Ho}, {Honma}, {Huang}, {Huang}, {Hughes}, {Ikeda}, {Impellizzeri}, {Inoue},
  {Issaoun}, {James}, {Jannuzi}, {Janssen}, {Jeter}, {Jiang},
  {Jim{\'e}nez-Rosales}, {Johnson}, {Jorstad}, {Joshi}, {Jung}, {Karami},
  {Karuppusamy}, {Kawashima}, {Keating}, {Kettenis}, {Kim}, {Kim}, {Kim},
  {Kim}, {Kino}, {Koay}, {Kocherlakota}, {Kofuji}, {Koch}, {Koyama}, {Kramer},
  {Kramer}, {Krichbaum}, {Kuo}, {Bella}, {Lauer}, {Lee}, {Lee}, {Leung},
  {Levis}, {Li}, {Lico}, {Lindahl}, {Lindqvist}, {Lisakov}, {Liu}, {Liu},
  {Liuzzo}, {Lo}, {Lobanov}, {Loinard}, {Lonsdale}, {Lu}, {Mao}, {Marchili},
  {Markoff}, {Marrone}, {Marscher}, {Mart{\'\i}-Vidal}, {Matsushita},
  {Matthews}, {Medeiros}, {Menten}, {Michalik}, {Mizuno}, {Mizuno}, {Moran},
  {Moriyama}, {Moscibrodzka}, {M{\"u}ller}, {Mus}, {Musoke}, {Myserlis},
  {Nadolski}, {Nagai}, {Nagar}, {Nakamura}, {Narayan}, {Narayanan},
  {Natarajan}, {Nathanail}, {Fuentes}, {Neilsen}, {Neri}, {Ni}, {Noutsos},
  {Nowak}, {Oh}, {Okino}, {Olivares}, {Ortiz-Le{\'o}n}, {Oyama}, {{\"O}zel},
  {Palumbo}, {Paraschos}, {Park}, {Parsons}, {Patel}, {Pen}, {Pesce},
  {Pi{\'e}tu}, {Plambeck}, {PopStefanija}, {Porth}, {P{\"o}tzl}, {Prather},
  {Preciado-L{\'o}pez}, {Psaltis}, {Pu}, {Ramakrishnan}, {Rao}, {Rawlings},
  {Raymond}, {Rezzolla}, {Ricarte}, {Ripperda}, {Roelofs}, {Rogers}, {Ros},
  {Romero-Ca{\~n}izales}, {Roshanineshat}, {Rottmann}, {Roy}, {Ruiz},
  {Ruszczyk}, {Rygl}, {S{\'a}nchez}, {S{\'a}nchez-Arg{\"u}elles},
  {S{\'a}nchez-Portal}, {Sasada}, {Satapathy}, {Savolainen}, {Schloerb},
  {Schonfeld}, {Schuster}, {Shao}, {Shen}, {Small}, {Sohn}, {SooHoo},
  {Souccar}, {Sun}, {Tazaki}, {Tetarenko}, {Tiede}, {Tilanus}, {Titus},
  {Torne}, {Traianou}, {Trent}, {Trippe}, {Turk}, {van Bemmel}, {van
  Langevelde}, {van Rossum}, {Vos}, {Wagner}, {Ward-Thompson}, {Wardle},
  {Weintroub}, {Wex}, {Wharton}, {Wielgus}, {Wiik}, {Witzel}, {Wondrak},
  {Wong}, {Wu}, {Yamaguchi}, {Yoon}, {Young}, {Young}, {Younsi}, {Yuan},
  {Yuan}, {Zensus}, {Zhang}, {Zhao}, {Zhao}, {Agurto}, {Allardi}, {Amestica},
  {Araneda}, {Arriagada}, {Berghuis}, {Bertarini}, {Berthold}, {Blanchard},
  {Brown}, {C{\'a}rdenas}, {Cantzler}, {Caro}, {Castillo-Dom{\'\i}nguez},
  {Chan}, {Chang}, {Chang}, {Chang}, {Chang}, {Chen}, {Chilson}, {Chuter},
  {Ciechanowicz}, {Colin-Beltran}, {Coulson}, {Crowley}, {Degenaar},
  {Dornbusch}, {Dur{\'a}n}, {Everett}, {Faber}, {Forster}, {Fuchs}, {Gale},
  {Geertsema}, {Gonz{\'a}lez}, {Graham}, {Gueth}, {Halverson}, {Han}, {Han},
  {Hasegawa}, {Hern{\'a}ndez-Rebollar}, {Herrera}, {Herrero-Illana},
  {Heyminck}, {Hirota}, {Hoge}, {Hostler Schimpf}, {Howie}, {Huang}, {Jiang},
  {Jinchi}, {John}, {Kimura}, {Klein}, {Kubo}, {Kuroda}, {Kwon}, {Lacasse},
  {Laing}, {Leitch}, {Li}, {Liu}, {Liu}, {Lin}, {Lu}, {Mac-Auliffe},
  {Martin-Cocher}, {Matulonis}, {Maute}, {Messias}, {Meyer-Zhao},
  {Monta{\~n}a}, {Montenegro-Montes}, {Montgomerie}, {Moreno Nolasco},
  {Muders}, {Nishioka}, {Norton}, {Nystrom}, {Ogawa}, {Olivares}, {Oshiro},
  {P{\'e}rez-Beaupuits}, {Parra}, {Phillips}, {Poirier}, {Pradel}, {Qiu},
  {Raffin}, {Rahlin}, {Ram{\'\i}rez}, {Ressler}, {Reynolds},
  {Rodr{\'\i}guez-Montoya}, {Saez-Madain}, {Santana}, {Shaw}, {Shirkey},
  {Silva}, {Snow}, {Sousa}, {Sridharan}, {Stahm}, {Stark}, {Test},
  {Torstensson}, {Venegas}, {Walther}, {Wei}, {White}, {Wieching}, {Wijnands},
  {Wouterloot}, {Yu}, {Yu (于威)}, \& {Zeballos}}]{EHTC_L12+22}
{Event Horizon Telescope Collaboration}, {Akiyama}, K., {Alberdi}, A., {et~al.}
  2022, \apjl, 930, L12

\bibitem[{{Fabian}(2012)}]{Fabian2012}
{Fabian}, A.~C. 2012, \araa, 50, 455

\bibitem[{{Falco} {et~al.}(1999){Falco}, {Kurtz}, {Geller}, {Huchra}, {Peters},
  {Berlind}, {Mink}, {Tokarz}, \& {Elwell}}]{Falco99}
{Falco}, E.~E., {Kurtz}, M.~J., {Geller}, M.~J., {et~al.} 1999, \pasp, 111, 438

\bibitem[{{Fischer} {et~al.}(2011){Fischer}, {Crenshaw}, {Kraemer}, {Schmitt},
  {Mushotsky}, \& {Dunn}}]{Fischer11}
{Fischer}, T.~C., {Crenshaw}, D.~M., {Kraemer}, S.~B., {et~al.} 2011, \apj,
  727, 71

\bibitem[{{Flagey} {et~al.}(2013){Flagey}, {Goldsmith}, {Lis}, {Gerin},
  {Neufeld}, {Sonnentrucker}, {De Luca}, {Godard}, {Goicoechea}, {Monje}, \&
  {Phillips}}]{Flagey+13}
{Flagey}, N., {Goldsmith}, P.~F., {Lis}, D.~C., {et~al.} 2013, \apj, 762, 11

\bibitem[{{Garc{\'\i}a-Bernete} {et~al.}(2022){Garc{\'\i}a-Bernete},
  {Rigopoulou}, {Alonso-Herrero}, {Pereira-Santaella}, {Roche}, \&
  {Kerkeni}}]{Garcia-Bernete+22}
{Garc{\'\i}a-Bernete}, I., {Rigopoulou}, D., {Alonso-Herrero}, A., {et~al.}
  2022, \mnras, 509, 4256

\bibitem[{{Gasman} {et~al.}(2022){Gasman}, {Argyriou}, {Sloan}, {Aringer},
  {{\'A}lvarez-M{\'a}rquez}, {Fox}, {Glasse}, {Glauser}, {Jones}, {Justtanont},
  {Kavanagh}, {Klaassen}, {Labiano}, {Larson}, {Law}, {Mueller}, {Nayak},
  {Noriega-Crespo}, {Patapis}, {Royer}, \& {Vandenbussche}}]{Gasman+2022}
{Gasman}, D., {Argyriou}, I., {Sloan}, G.~C., {et~al.} 2022, arXiv e-prints,
  arXiv:2212.03596

\bibitem[{{G{\'a}sp{\'a}r} {et~al.}(2021){G{\'a}sp{\'a}r}, {Rieke}, {Guillard},
  {Dicken}, {Gastaud}, {Alberts}, {Morrison}, {Ressler}, {Argyriou}, \&
  {Glasse}}]{Gaspar+21}
{G{\'a}sp{\'a}r}, A., {Rieke}, G.~H., {Guillard}, P., {et~al.} 2021, \pasp,
  133, 014504

\bibitem[{{Ghez} {et~al.}(2008){Ghez}, {Salim}, {Weinberg}, {Lu}, {Do}, {Dunn},
  {Matthews}, {Morris}, {Yelda}, {Becklin}, {Kremenek}, {Milosavljevic}, \&
  {Naiman}}]{Ghez2008}
{Ghez}, A.~M., {Salim}, S., {Weinberg}, N.~N., {et~al.} 2008, \apj, 689, 1044

\bibitem[{{Gioia} {et~al.}(2003){Gioia}, {Henry}, {Mullis}, {B{\"o}hringer},
  {Briel}, {Voges}, \& {Huchra}}]{Gioia03}
{Gioia}, I.~M., {Henry}, J.~P., {Mullis}, C.~R., {et~al.} 2003, \apjs, 149, 29

\bibitem[{{Glasse} {et~al.}(2015){Glasse}, {Rieke}, {Bauwens},
  {Garc{\'\i}a-Mar{\'\i}n}, {Ressler}, {Rost}, {Tikkanen}, {Vandenbussche}, \&
  {Wright}}]{Glasse+15}
{Glasse}, A., {Rieke}, G.~H., {Bauwens}, E., {et~al.} 2015, \pasp, 127, 686

\bibitem[{{Guillard} {et~al.}(2015){Guillard}, {Boulanger}, {Lehnert},
  {Appleton}, \& {Pineau des For{\^e}ts}}]{Guillard+15}
{Guillard}, P., {Boulanger}, F., {Lehnert}, M.~D., {Appleton}, P.~N., \&
  {Pineau des For{\^e}ts}, G. 2015, in SF2A-2015: Proceedings of the Annual
  meeting of the French Society of Astronomy and Astrophysics, 81--85

\bibitem[{{Guillard} {et~al.}(2012){Guillard}, {Ogle}, {Emonts}, {Appleton},
  {Morganti}, {Tadhunter}, {Oosterloo}, {Evans}, \& {Evans}}]{Guillard+12}
{Guillard}, P., {Ogle}, P.~M., {Emonts}, B.~H.~C., {et~al.} 2012, \apj, 747, 95

\bibitem[{{Habart} {et~al.}(2005){Habart}, {Walmsley}, {Verstraete}, {Cazaux},
  {Maiolino}, {Cox}, {Boulanger}, \& {Pineau des For{\^e}ts}}]{Habart+05}
{Habart}, E., {Walmsley}, M., {Verstraete}, L., {et~al.} 2005, \ssr, 119, 71

\bibitem[{{Harrison} {et~al.}(2016){Harrison}, {Alexander}, {Mullaney},
  {Stott}, {Swinbank}, {Arumugam}, {Bauer}, {Bower}, {Bunker}, \&
  {Sharples}}]{Harrison+16}
{Harrison}, C.~M., {Alexander}, D.~M., {Mullaney}, J.~R., {et~al.} 2016,
  \mnras, 456, 1195

\bibitem[{{Harrison} {et~al.}(2013){Harrison}, {Craig}, {Christensen},
  {Hailey}, {Zhang}, {Boggs}, {Stern}, {Cook}, {Forster}, {Giommi},
  {Grefenstette}, {Kim}, {Kitaguchi}, {Koglin}, {Madsen}, {Mao}, {Miyasaka},
  {Mori}, {Perri}, {Pivovaroff}, {Puccetti}, {Rana}, {Westergaard}, {Willis},
  {Zoglauer}, {An}, {Bachetti}, {Barri{\`e}re}, {Bellm}, {Bhalerao},
  {Brejnholt}, {Fuerst}, {Liebe}, {Markwardt}, {Nynka}, {Vogel}, {Walton},
  {Wik}, {Alexander}, {Cominsky}, {Hornschemeier}, {Hornstrup}, {Kaspi},
  {Madejski}, {Matt}, {Molendi}, {Smith}, {Tomsick}, {Ajello}, {Ballantyne},
  {Balokovi{\'c}}, {Barret}, {Bauer}, {Blandford}, {Brandt}, {Brenneman},
  {Chiang}, {Chakrabarty}, {Chenevez}, {Comastri}, {Dufour}, {Elvis}, {Fabian},
  {Farrah}, {Fryer}, {Gotthelf}, {Grindlay}, {Helfand}, {Krivonos}, {Meier},
  {Miller}, {Natalucci}, {Ogle}, {Ofek}, {Ptak}, {Reynolds}, {Rigby},
  {Tagliaferri}, {Thorsett}, {Treister}, \& {Urry}}]{Harrison+13}
{Harrison}, F.~A., {Craig}, W.~W., {Christensen}, F.~E., {et~al.} 2013, \apj,
  770, 103

\bibitem[{{Jansen} {et~al.}(2001){Jansen}, {Lumb}, {Altieri}, {Clavel}, {Ehle},
  {Erd}, {Gabriel}, {Guainazzi}, {Gondoin}, {Much}, {Munoz}, {Santos},
  {Schartel}, {Texier}, \& {Vacanti}}]{Jansen+01}
{Jansen}, F., {Lumb}, D., {Altieri}, B., {et~al.} 2001, \aap, 365, L1

\bibitem[{{Jarrett} {et~al.}(2011){Jarrett}, {Cohen}, {Masci}, {Wright},
  {Stern}, {Benford}, {Blain}, {Carey}, {Cutri}, {Eisenhardt}, {Lonsdale},
  {Mainzer}, {Marsh}, {Padgett}, {Petty}, {Ressler}, {Skrutskie}, {Stanford},
  {Surace}, {Tsai}, {Wheelock}, \& {Yan}}]{Jarrett11}
{Jarrett}, T.~H., {Cohen}, M., {Masci}, F., {et~al.} 2011, \apj, 735, 112

\bibitem[{{Jones} {et~al.}(2023){Jones}, {{\'A}lvarez-M{\'a}rquez}, {Sloan},
  {Kavanagh}, {Argyriou}, {Labiano}, {Law}, {Patapis}, {Mueller}, {Larson},
  {Bright}, {Klaassen}, {Fox}, {Gasman}, {Geers}, {Glauser}, {Guillard},
  {Nayak}, {Noriega-Crespo}, {Ressler}, {Sargent}, {Temim}, {Vandenbussche}, \&
  {Garc{\'\i}a Mar{\'\i}n}}]{Jones+23}
{Jones}, O.~C., {{\'A}lvarez-M{\'a}rquez}, J., {Sloan}, G.~C., {et~al.} 2023,
  arXiv e-prints, arXiv:2301.13233

\bibitem[{{King} \& {Pounds}(2015)}]{King+15}
{King}, A. \& {Pounds}, K. 2015, \araa, 53, 115

\bibitem[{{Labiano} {et~al.}(2021){Labiano}, {Argyriou},
  {{\'A}lvarez-M{\'a}rquez}, {Glasse}, {Glauser}, {Patapis}, {Law}, {Brandl},
  {Justtanont}, {Lahuis}, {Mart{\'\i}nez-Galarza}, {Mueller}, {Noriega-Crespo},
  {Royer}, {Shaughnessy}, \& {Vandenbussche}}]{Labiano21}
{Labiano}, A., {Argyriou}, I., {{\'A}lvarez-M{\'a}rquez}, J., {et~al.} 2021,
  \aap, 656, A57

\bibitem[{{Labiano} {et~al.}(2016){Labiano}, {Azzollini}, {Bailey}, {Beard},
  {Dicken}, {Garc{\'\i}a-Mar{\'\i}n}, {Geers}, {Glasse}, {Glauser}, {Gordon},
  {Justtanont}, {Klaassen}, {Lahuis}, {Law}, {Morrison}, {M{\"u}ller}, {Rieke},
  {Vandenbussche}, \& {Wright}}]{MRSpipeline}
{Labiano}, A., {Azzollini}, R., {Bailey}, J., {et~al.} 2016, in SPIE Conf.
  Series, Vol. 9910,

\bibitem[{{Markwardt}(2009)}]{Markwardt2009}
{Markwardt}, C.~B. 2009, in Astronomical Society of the Pacific Conference
  Series, Vol. 411, Astronomical Data Analysis Software and Systems XVIII, ed.
  D.~A. {Bohlender}, D.~{Durand}, \& P.~{Dowler}, 251

\bibitem[{{Moran} {et~al.}(1996){Moran}, {Halpern}, \& {Helfand}}]{Moran1996}
{Moran}, E.~C., {Halpern}, J.~P., \& {Helfand}, D.~J. 1996, \apjs, 106, 341

\bibitem[{{Neufeld} {et~al.}(1998){Neufeld}, {Melnick}, \&
  {Harwit}}]{Neufeld+98}
{Neufeld}, D.~A., {Melnick}, G.~J., \& {Harwit}, M. 1998, \apjl, 506, L75

\bibitem[{{Ogle} {et~al.}(2010){Ogle}, {Boulanger}, {Guillard}, {Evans},
  {Antonucci}, {Appleton}, {Nesvadba}, \& {Leipski}}]{Ogle+10}
{Ogle}, P., {Boulanger}, F., {Guillard}, P., {et~al.} 2010, \apj, 724, 1193

\bibitem[{{Paturel} {et~al.}(2002){Paturel}, {Dubois}, {Petit}, \&
  {Woelfel}}]{2002LEDA}
{Paturel}, G., {Dubois}, P., {Petit}, C., \& {Woelfel}, F. 2002, LEDA, 0

\bibitem[{{Pereira-Santaella} {et~al.}(2022){Pereira-Santaella},
  {{\'A}lvarez-M{\'a}rquez}, {Garc{\'\i}a-Bernete}, {Labiano}, {Colina},
  {Alonso-Herrero}, {Bellocchi}, {Garc{\'\i}a-Burillo}, {H{\"o}nig}, {Ramos
  Almeida}, \& {Rosario}}]{Pereira22}
{Pereira-Santaella}, M., {{\'A}lvarez-M{\'a}rquez}, J., {Garc{\'\i}a-Bernete},
  I., {et~al.} 2022, arXiv e-prints, arXiv:2208.04835

\bibitem[{{Pereira-Santaella} {et~al.}(2018){Pereira-Santaella}, {Colina},
  {Garc{\'\i}a-Burillo}, {Combes}, {Emonts}, {Aalto}, {Alonso-Herrero},
  {Arribas}, {Henkel}, {Labiano}, {Muller}, {Piqueras L{\'o}pez}, {Rigopoulou},
  \& {van der Werf}}]{Pereira-Santaella+18}
{Pereira-Santaella}, M., {Colina}, L., {Garc{\'\i}a-Burillo}, S., {et~al.}
  2018, \aap, 616, A171

\bibitem[{{Pereira-Santaella} {et~al.}(2010){Pereira-Santaella},
  {Diamond-Stanic}, {Alonso-Herrero}, \& {Rieke}}]{Pereira-Santaella+10}
{Pereira-Santaella}, M., {Diamond-Stanic}, A.~M., {Alonso-Herrero}, A., \&
  {Rieke}, G.~H. 2010, \apj, 725, 2270

\bibitem[{{Petric} {et~al.}(2018){Petric}, {Armus}, {Flagey}, {Guillard},
  {Howell}, {Inami}, {Charmandaris}, {Evans}, {Stierwalt}, {Diaz-Santos}, {Lu},
  {Spoon}, {Mazzarella}, {Appleton}, {Chan}, {Chu}, {Hand}, {Privon},
  {Sanders}, {Surace}, {Xu}, \& {Zhao}}]{Petric+18}
{Petric}, A.~O., {Armus}, L., {Flagey}, N., {et~al.} 2018, \aj, 156, 295

\bibitem[{{Planck Collaboration} {et~al.}(2016){Planck Collaboration}, {Ade},
  {Aghanim}, {Arnaud}, {Ashdown}, {Aumont}, {Baccigalupi}, {Banday},
  {Barreiro}, {Bartlett}, {Bartolo}, {Battaner}, {Battye}, {Benabed},
  {Beno{\^\i}t}, {Benoit-L{\'e}vy}, {Bernard}, {Bersanelli}, {Bielewicz},
  {Bock}, {Bonaldi}, {Bonavera}, {Bond}, {Borrill}, {Bouchet}, {Boulanger},
  {Bucher}, {Burigana}, {Butler}, {Calabrese}, {Cardoso}, {Catalano},
  {Challinor}, {Chamballu}, {Chary}, {Chiang}, {Chluba}, {Christensen},
  {Church}, {Clements}, {Colombi}, {Colombo}, {Combet}, {Coulais}, {Crill},
  {Curto}, {Cuttaia}, {Danese}, {Davies}, {Davis}, {de Bernardis}, {de Rosa},
  {de Zotti}, {Delabrouille}, {D{\'e}sert}, {Di Valentino}, {Dickinson},
  {Diego}, {Dolag}, {Dole}, {Donzelli}, {Dor{\'e}}, {Douspis}, {Ducout},
  {Dunkley}, {Dupac}, {Efstathiou}, {Elsner}, {En{\ss}lin}, {Eriksen},
  {Farhang}, {Fergusson}, {Finelli}, {Forni}, {Frailis}, {Fraisse},
  {Franceschi}, {Frejsel}, {Galeotta}, {Galli}, {Ganga}, {Gauthier}, {Gerbino},
  {Ghosh}, {Giard}, {Giraud-H{\'e}raud}, {Giusarma}, {Gjerl{\o}w},
  {Gonz{\'a}lez-Nuevo}, {G{\'o}rski}, {Gratton}, {Gregorio}, {Gruppuso},
  {Gudmundsson}, {Hamann}, {Hansen}, {Hanson}, {Harrison}, {Helou},
  {Henrot-Versill{\'e}}, {Hern{\'a}ndez-Monteagudo}, {Herranz}, {Hildebrand t},
  {Hivon}, {Hobson}, {Holmes}, {Hornstrup}, {Hovest}, {Huang}, {Huffenberger},
  {Hurier}, {Jaffe}, {Jaffe}, {Jones}, {Juvela}, {Keih{\"a}nen}, {Keskitalo},
  {Kisner}, {Kneissl}, {Knoche}, {Knox}, {Kunz}, {Kurki-Suonio}, {Lagache},
  {L{\"a}hteenm{\"a}ki}, {Lamarre}, {Lasenby}, {Lattanzi}, {Lawrence}, {Leahy},
  {Leonardi}, {Lesgourgues}, {Levrier}, {Lewis}, {Liguori}, {Lilje},
  {Linden-V{\o}rnle}, {L{\'o}pez-Caniego}, {Lubin}, {Mac{\'\i}as-P{\'e}rez},
  {Maggio}, {Maino}, {Mandolesi}, {Mangilli}, {Marchini}, {Maris}, {Martin},
  {Martinelli}, {Mart{\'\i}nez-Gonz{\'a}lez}, {Masi}, {Matarrese}, {McGehee},
  {Meinhold}, {Melchiorri}, {Melin}, {Mendes}, {Mennella}, {Migliaccio},
  {Millea}, {Mitra}, {Miville-Desch{\^e}nes}, {Moneti}, {Montier}, {Morgante},
  {Mortlock}, {Moss}, {Munshi}, {Murphy}, {Naselsky}, {Nati}, {Natoli},
  {Netterfield}, {N{\o}rgaard-Nielsen}, {Noviello}, {Novikov}, {Novikov},
  {Oxborrow}, {Paci}, {Pagano}, {Pajot}, {Paladini}, {Paoletti}, {Partridge},
  {Pasian}, {Patanchon}, {Pearson}, {Perdereau}, {Perotto}, {Perrotta},
  {Pettorino}, {Piacentini}, {Piat}, {Pierpaoli}, {Pietrobon}, {Plaszczynski},
  {Pointecouteau}, {Polenta}, {Popa}, {Pratt}, {Pr{\'e}zeau}, {Prunet},
  {Puget}, {Rachen}, {Reach}, {Rebolo}, {Reinecke}, {Remazeilles}, {Renault},
  {Renzi}, {Ristorcelli}, {Rocha}, {Rosset}, {Rossetti}, {Roudier},
  {Rouill{\'e} d'Orfeuil}, {Rowan-Robinson}, {Rubi{\~n}o-Mart{\'\i}n},
  {Rusholme}, {Said}, {Salvatelli}, {Salvati}, {Sandri}, {Santos},
  {Savelainen}, {Savini}, {Scott}, {Seiffert}, {Serra}, {Shellard}, {Spencer},
  {Spinelli}, {Stolyarov}, {Stompor}, {Sudiwala}, {Sunyaev}, {Sutton},
  {Suur-Uski}, {Sygnet}, {Tauber}, {Terenzi}, {Toffolatti}, {Tomasi},
  {Tristram}, {Trombetti}, {Tucci}, {Tuovinen}, {T{\"u}rler}, {Umana},
  {Valenziano}, {Valiviita}, {Van Tent}, {Vielva}, {Villa}, {Wade}, {Wandelt},
  {Wehus}, {White}, {White}, {Wilkinson}, {Yvon}, {Zacchei}, \&
  {Zonca}}]{Cosmology2016}
{Planck Collaboration}, {Ade}, P.~A.~R., {Aghanim}, N., {et~al.} 2016, \aap,
  594, A13

\bibitem[{{Ressler} {et~al.}(2015){Ressler}, {Sukhatme}, {Franklin}, {Mahoney},
  {Thelen}, {Bouchet}, {Colbert}, {Cracraft}, {Dicken}, {Gastaud}, {Goodson},
  {Eccleston}, {Moreau}, {Rieke}, \& {Schneider}}]{Ressler+15}
{Ressler}, M.~E., {Sukhatme}, K.~G., {Franklin}, B.~R., {et~al.} 2015, \pasp,
  127, 675

\bibitem[{{Ricci} {et~al.}(2015){Ricci}, {Ueda}, {Koss}, {Trakhtenbrot},
  {Bauer}, \& {Gandhi}}]{Ricci2015}
{Ricci}, C., {Ueda}, Y., {Koss}, M.~J., {et~al.} 2015, \apjl, 815, L13

\bibitem[{{Rieke} {et~al.}(2015{\natexlab{a}}){Rieke}, {Ressler}, {Morrison},
  {Bergeron}, {Bouchet}, {Garc{\'\i}a-Mar{\'\i}n}, {Greene}, {Regan},
  {Sukhatme}, \& {Walker}}]{Rieke+15}
{Rieke}, G.~H., {Ressler}, M.~E., {Morrison}, J.~E., {et~al.}
  2015{\natexlab{a}}, \pasp, 127, 665

\bibitem[{{Rieke} {et~al.}(2015{\natexlab{b}}){Rieke}, {Wright}, {B{\"o}ker},
  {Bouwman}, {Colina}, {Glasse}, {Gordon}, {Greene}, {G{\"u}del}, {Henning},
  {Justtanont}, {Lagage}, {Meixner}, {N{\o}rgaard-Nielsen}, {Ray}, {Ressler},
  {van Dishoeck}, \& {Waelkens}}]{miri_pasp_1}
{Rieke}, G.~H., {Wright}, G.~S., {B{\"o}ker}, T., {et~al.} 2015{\natexlab{b}},
  PASP, 127, 584

\bibitem[{{Rigby} {et~al.}(2022){Rigby}, {Perrin}, {McElwain}, {Kimble},
  {Friedman}, {Lallo}, {Doyon}, {Feinberg}, {Ferruit}, {Glasse}, {Rieke},
  {Rieke}, {Wright}, {Willott}, {Colon}, {Milam}, {Neff}, {Stark}, {Valenti},
  {Abell}, {Abney}, {Abul-Huda}, {Acton}, {Adams}, {Adler}, {Aguilar}, {Ahmed},
  {Albert}, {Alberts}, {Aldridge}, {Allen}, {Altenburg}, {Alves de Oliveira},
  {Anderson}, {Anderson}, {Anderson}, {Argyriou}, {Armstrong}, {Arribas},
  {Artigau}, {Arvai}, {Atkinson}, {Bacon}, {Bair}, {Banks}, {Barrientes},
  {Barringer}, {Bartosik}, {Bast}, {Baudoz}, {Beatty}, {Bechtold}, {Beck},
  {Bergeron}, {Bergkoetter}, {Bhatawdekar}, {Birkmann}, {Blazek}, {Blome},
  {Boccaletti}, {Boeker}, {Boia}, {Bonaventura}, {Bond}, {Bosley}, {Boucarut},
  {Bourque}, {Bouwman}, {Bower}, {Bowers}, {Boyer}, {Brady}, {Braun}, {Breda},
  {Bresnahan}, {Bright}, {Britt}, {Bromenschenkel}, {Brooks}, {Brooks},
  {Brown}, {Brown}, {Brown}, {Bunker}, {Burger}, {Bushouse}, {Cale}, {Cameron},
  {Cameron}, {Canipe}, {Caplinger}, {Caputo}, {Carey}, {Carniani},
  {Carrasquilla}, {Carruthers}, {Case}, {Chance}, {Chapman}, {Charlot},
  {Charlow}, {Chayer}, {Chen}, {Cherinka}, {Chichester}, {Chilton}, {Chonis},
  {Clark}, {Clark}, {Coe}, {Coleman}, {Comber}, {Comeau}, {Connolly}, {Cooper},
  {Cooper}, {Coppock}, {Correnti}, {Cossou}, {Coulais}, {Coyle}, {Cracraft},
  {Curti}, {Cuturic}, {Davis}, {Davis}, {Dean}, {DeLisa}, {deMeester},
  {Dencheva}, {Dencheva}, {DePasquale}, {Deschenes}, {Hunor Detre}, {Diaz},
  {Dicken}, {DiFelice}, {Dillman}, {Dixon}, {Doggett}, {Donaldson}, {Douglas},
  {DuPrie}, {Dupuis}, {Durning}, {Easmin}, {Eck}, {Edeani}, {Egami},
  {Ehrenwinkler}, {Eisenhamer}, {Eisenhower}, {Elie}, {Elliott}, {Elliott},
  {Ellis}, {Engesser}, {Espinoza}, {Etienne}, {Etxaluze}, {Falini}, {Feeney},
  {Ferry}, {Filippazzo}, {Fincham}, {Fix}, {Flagey}, {Florian}, {Flynn},
  {Fontanella}, {Ford}, {Forshay}, {Fox}, {Franz}, {Fu}, {Fullerton}, {Galkin},
  {Galyer}, {Garcia Marin}, {Gardner}, {Gardner}, {Garland}, {Gasman},
  {Gaspar}, {Gaudreau}, {Gauthier}, {Geers}, {Geithner}, {Gennaro}, {Giardino},
  {Girard}, {Giuliano}, {Glassmire}, {Glauser}, {Glazer}, {Godfrey},
  {Golimowski}, {Gollnitz}, {Gong}, {Gonzaga}, {Gordon}, {Gordon},
  {Goudfrooij}, {Greene}, {Greenhouse}, {Grimaldi}, {Groebner}, {Grundy},
  {Guillard}, {Gutman}, {Ha}, {Haderlein}, {Hagedorn}, {Hainline}, {Haley},
  {Hami}, {Hamilton}, {Hammel}, {Hansen}, {Harkins}, {Harr}, {Hart}, {Hart},
  {Hartig}, {Hashimoto}, {Haskins}, {Hathaway}, {Havey}, {Hayden}, {Hecht},
  {Heller-Boyer}, {Henry}, {Hermann}, {Hernandez}, {Hesman}, {Hicks},
  {Hilbert}, {Hines}, {Hoffman}, {Holfeltz}, {Holler}, {Hoppa}, {Hott},
  {Howard}, {Hunter}, {Hunter}, {Hurst}, {Husemann}, {Hustak}, {Ilinca Ignat},
  {Irish}, {Jackson}, {Jahromi}, {Jakobsen}, {James}, {James}, {Januszewski},
  {Jenkins}, {Jirdeh}, {Johnson}, {Johnson}, {Jones}, {Jones}, {Jones},
  {Jones}, {Jordan}, {Jordan}, {Jurczyk}, {Jurling}, {Kaleida}, {Kalmanson},
  {Kammerer}, {Kang}, {Kao}, {Karakla}, {Kavanagh}, {Kelly}, {Kendrew},
  {Kennedy}, {Kenny}, {Keski-kuha}, {Keyes}, {Kidwell}, {Kinzel}, {Kirk},
  {Kirkpatrick}, {Kirshenblat}, {Klaassen}, {Knapp}, {Knight}, {Knollenberg},
  {Koehler}, {Koekemoer}, {Kovacs}, {Kulp}, {Kumari}, {Kyprianou}, {La Massa},
  {Labador}, {Labiano Ortega}, {Lagage}, {Lajoie}, {Lallo}, {Lam}, {Lamb},
  {Lambros}, {Lampenfield}, {Langston}, {Larson}, {Law}, {Lawrence}, {Lee},
  {Leisenring}, {Lepo}, {Leveille}, {Levenson}, {Levine}, {Levy}, {Lewis},
  {Lewis}, {Libralato}, {Lightsey}, {Link}, {Liu}, {Lo}, {Lockwood}, {Logue},
  {Long}, {Long}, {Loomis}, {Lopez-Caniego}, {Alvarez}, {Love-Pruitt}, {Lucy},
  {Luetzgendorf}, {Maghami}, {Maiolino}, {Major}, {Malla}, {Malumuth},
  {Manjavacas}, {Mannfolk}, {Marrione}, {Marston}, {Martel}, {Maschmann},
  {Masci}, {Masciarelli}, {Maszkiewicz}, {Mather}, {McKenzie}, {McLean},
  {McMaster}, {Melbourne}, {Mel{\'e}ndez}, {Menzel}, {Merz}, {Meyett}, {Meza},
  {Miskey}, {Misselt}, {Moller}, {Morrison}, {Morse}, {Moseley}, {Mosier},
  {Mountain}, {Mueckay}, {Mueller}, {Mullally}, {Murphy}, {Murray}, {Murray},
  {Muzerolle}, {Mycroft}, {Myers}, {Myrick}, {Nanavati}, {Nance}, {Nayak},
  {Naylor}, {Nelan}, {Nickson}, {Nielson}, {Nieto-Santisteban}, {Nikolov},
  {Noriega-Crespo}, {O'Shaughnessy}, {O'Sullivan}, {Ochs}, {Ogle}, {Oleszczuk},
  {Olmsted}, {Osborne}, {Ottens}, {Owens}, {Pacifici}, {Pagan}, {Page},
  {Parrish}, {Patapis}, {Pauly}, {Pavlovsky}, {Pedder}, {Peek},
  {Pena-Guerrero}, {Pennanen}, {Perez}, {Perna}, {Perriello}, {Phillips},
  {Pietraszkiewicz}, {Pinaud}, {Pirzkal}, {Pitman}, {Piwowar}, {Platais},
  {Player}, {Plesha}, {Pollizi}, {Polster}, {Pontoppidan}, {Porterfield},
  {Proffitt}, {Pueyo}, {Pulliam}, {Quirt}, {Quispe Neira}, {Ramos Alarcon},
  {Ramsay}, {Rapp}, {Rapp}, {Rauscher}, {Ravindranath}, {Rawle}, {Regan},
  {Reichard}, {Reis}, {Ressler}, {Rest}, {Reynolds}, {Rhue}, {Richon},
  {Rickman}, {Ridgaway}, {Ritchie}, {Rix}, {Robberto}, {Robinson}, {Robinson},
  {Robinson}, {Rock}, {Rodriguez}, {Rodriguez Del Pino}, {Roellig}, {Rohrbach},
  {Roman}, {Romelfanger}, {Rose}, {Roteliuk}, {Roth}, {Rothwell}, {Rowlands},
  {Roy}, {Royer}, {Royle}, {Rui}, {Rumler}, {Runnels}, {Russ}, {Rustamkulov},
  {Ryden}, {Ryer}, {Sabata}, {Sabatke}, {Sabbi}, {Samuelson}, {Sappington},
  {Sargent}, {Sauer}, {Scheithauer}, {Schlawin}, {Schlitz}, {Schmitz},
  {Schneider}, {Schreiber}, {Schulze}, {Schwab}, {Scott}, {Sembach},
  {Shaughnessy}, {Shaw}, {Shawger}, {Shay}, {Sheehan}, {Shen}, {Sherman},
  {Shiao}, {Shih}, {Shivaei}, {Sienkiewicz}, {Sing}, {Sirianni},
  {Sivaramakrishnan}, {Skipper}, {Sloan}, {Slocum}, {Slowinski}, {Smith},
  {Smith}, {Smith}, {Smith}, {Snyder}, {Soh}, {Sohn}, {Soto}, {Spencer},
  {Stallcup}, {Stansberry}, {Starr}, {Starr}, {Stewart}, {Stiavelli},
  {Straughn}, {Strickland}, {Stys}, {Summers}, {Sun}, {Sunnquist}, {Swade},
  {Swam}, {Swaters}, {Swoish}, {Taylor}, {Taylor}, {Te Plate}, {Tea}, {Teague},
  {Telfer}, {Temim}, {Thatte}, {Thompson}, {Thompson}, {Thomson}, {Tikkanen},
  {Tippet}, {Todd}, {Toolan}, {Tran}, {Trejo}, {Truong}, {Tsukamoto},
  {Tustain}, {Tyra}, {Ubeda}, {Underwood}, {Uzzo}, {Van Campen}, {Vandal},
  {Vandenbussche}, {Vila}, {Volk}, {Wahlgren}, {Waldman}, {Walker}, {Wander},
  {Warfield}, {Warner}, {Wasiak}, {Watkins}, {Weilert}, {Weiser}, {Weiss},
  {Weissman}, {Welty}, {West}, {Wheate}, {Wheatley}, {Wheeler}, {White},
  {Whiteaker}, {Whitehouse}, {Whiteleather}, {Whitman}, {Williams}, {Willmer},
  {Willoughby}, {Wilson}, {Wirth}, {Wislowski}, {Wolf}, {Wolfe}, {Wolff},
  {Workman}, {Wright}, {Wu}, {Wu}, {Wymer}, {Yates}, {Yates}, {Yeager},
  {Yerger}, {Yoon}, {Young}, {Yu}, {Zak}, {Zeidler}, {Zhou}, {Zielinski},
  {Zincke}, \& {Zonak}}]{Rigby22}
{Rigby}, J., {Perrin}, M., {McElwain}, M., {et~al.} 2022, arXiv e-prints,
  arXiv:2207.05632

\bibitem[{{Rodr{\'\i}guez-Ardila} {et~al.}(2006){Rodr{\'\i}guez-Ardila},
  {Prieto}, {Viegas}, \& {Gruenwald}}]{Rodriguez-Ardila+06}
{Rodr{\'\i}guez-Ardila}, A., {Prieto}, M.~A., {Viegas}, S., \& {Gruenwald}, R.
  2006, \apj, 653, 1098

\bibitem[{{Shu} {et~al.}(2007){Shu}, {Wang}, {Jiang}, {Fan}, \& {Wang}}]{Shu07}
{Shu}, X.~W., {Wang}, J.~X., {Jiang}, P., {Fan}, L.~L., \& {Wang}, T.~G. 2007,
  \apj, 657, 167

\bibitem[{{Spilker} {et~al.}(2020){Spilker}, {Aravena}, {Phadke},
  {B{\'e}thermin}, {Chapman}, {Dong}, {Gonzalez}, {Hayward}, {Hezaveh},
  {Litke}, {Malkan}, {Marrone}, {Narayanan}, {Reuter}, {Vieira}, \&
  {Wei{\ss}}}]{Spilker+20}
{Spilker}, J.~S., {Aravena}, M., {Phadke}, K.~A., {et~al.} 2020, \apj, 905, 86

\bibitem[{{Togi} \& {Smith}(2016)}]{Togi+16}
{Togi}, A. \& {Smith}, J.~D.~T. 2016, \apj, 830, 18

\bibitem[{{Torres-Alb{\`a}} {et~al.}(2021){Torres-Alb{\`a}}, {Marchesi},
  {Zhao}, {Ajello}, {Silver}, {Ananna}, {Balokovi{\'c}}, {Boorman}, {Comastri},
  {Gilli}, {Lanzuisi}, {Murphy}, {Urry}, \& {Vignali}}]{Torres-Alba_21}
{Torres-Alb{\`a}}, N., {Marchesi}, S., {Zhao}, X., {et~al.} 2021, \apj, 922,
  252

\bibitem[{{Tran}(2001)}]{Tran+01}
{Tran}, H.~D. 2001, \apjl, 554, L19

\bibitem[{{Veilleux} {et~al.}(2005){Veilleux}, {Cecil}, \&
  {Bland-Hawthorn}}]{Veilleux+05}
{Veilleux}, S., {Cecil}, G., \& {Bland-Hawthorn}, J. 2005, \araa, 43, 769

\bibitem[{{Veilleux} {et~al.}(2020){Veilleux}, {Maiolino}, {Bolatto}, \&
  {Aalto}}]{Veilleux+20}
{Veilleux}, S., {Maiolino}, R., {Bolatto}, A.~D., \& {Aalto}, S. 2020, \aapr,
  28, 2

\bibitem[{{Veilleux} {et~al.}(2013){Veilleux}, {Mel{\'e}ndez}, {Sturm},
  {Gracia-Carpio}, {Fischer}, {Gonz{\'a}lez-Alfonso}, {Contursi}, {Lutz},
  {Poglitsch}, {Davies}, {Genzel}, {Tacconi}, {de Jong}, {Sternberg}, {Netzer},
  {Hailey-Dunsheath}, {Verma}, {Rupke}, {Maiolino}, {Teng}, \&
  {Polisensky}}]{Veilleux+13}
{Veilleux}, S., {Mel{\'e}ndez}, M., {Sturm}, E., {et~al.} 2013, \apj, 776, 27

\bibitem[{{Wakelam} {et~al.}(2017){Wakelam}, {Bron}, {Cazaux}, {Dulieu}, {Gry},
  {Guillard}, {Habart}, {Hornek{\ae}r}, {Morisset}, {Nyman}, {Pirronello},
  {Price}, {Valdivia}, {Vidali}, \& {Watanabe}}]{Wakelam+17}
{Wakelam}, V., {Bron}, E., {Cazaux}, S., {et~al.} 2017, Molecular Astrophysics,
  9, 1

\bibitem[{Wells {et~al.}(2015)Wells, Pel, Glasse, Wright, Aitink-Kroes,
  Azzollini, Beard, Brandl, Gallie, Geers, \& et~al.}]{Wells15}
Wells, M., Pel, J.-W., Glasse, A., {et~al.} 2015, PASP, 127, 646–664

\bibitem[{{Wilgenbus} {et~al.}(2000){Wilgenbus}, {Cabrit}, {Pineau des
  For{\^e}ts}, \& {Flower}}]{Wilgenbus+00}
{Wilgenbus}, D., {Cabrit}, S., {Pineau des For{\^e}ts}, G., \& {Flower}, D.~R.
  2000, \aap, 356, 1010

\bibitem[{{Wright}(2006)}]{CosmoCalc}
{Wright}, E.~L. 2006, \pasp, 118, 1711

\bibitem[{{Wright} {et~al.}(2015){Wright}, {Wright}, {Goodson}, {Rieke},
  {Aitink-Kroes}, {Amiaux}, {Aricha-Yanguas}, {Azzollini}, {Banks},
  {Barrado-Navascues}, {Belenguer-Davila}, {Bloemmart}, {Bouchet}, {Brandl},
  {Colina}, {Detre}, {Diaz-Catala}, {Eccleston}, {Friedman},
  {Garc{\'\i}a-Mar{\'\i}n}, {G{\"u}del}, {Glasse}, {Glauser}, {Greene},
  {Groezinger}, {Grundy}, {Hastings}, {Henning}, {Hofferbert}, {Hunter},
  {Jessen}, {Justtanont}, {Karnik}, {Khorrami}, {Krause}, {Labiano}, {Lagage},
  {Langer}, {Lemke}, {Lim}, {Lorenzo-Alvarez}, {Mazy}, {McGowan}, {Meixner},
  {Morris}, {Morrison}, {M{\"u}ller}, {rgaard-Nielson}, {Olofsson},
  {O'Sullivan}, {Pel}, {Penanen}, {Petach}, {Pye}, {Ray}, {Renotte}, {Renouf},
  {Ressler}, {Samara-Ratna}, {Scheithauer}, {Schneider}, {Shaughnessy},
  {Stevenson}, {Sukhatme}, {Swinyard}, {Sykes}, {Thatcher}, {Tikkanen}, {van
  Dishoeck}, {Waelkens}, {Walker}, {Wells}, \& {Zhender}}]{Wright+15}
{Wright}, G.~S., {Wright}, D., {Goodson}, G.~B., {et~al.} 2015, \pasp, 127, 595

\end{thebibliography}

\end{document}